\documentclass[12pt]{article}

\usepackage{tikz}
\usetikzlibrary{decorations.pathreplacing}
\usepackage{amsmath}
\usepackage{amssymb}
\usepackage{graphicx}
\usepackage{subcaption}
\usepackage{amsfonts}
\usepackage{marginnote}
\usepackage{float}
\usepackage[utf8]{inputenc}
\usepackage[T1]{fontenc}
\usepackage{hyperref}
\usepackage{cleveref}
\usepackage{color}
\usepackage{comment}

\providecommand{\U}[1]{\protect\rule{.1in}{.1in}}
\newtheorem{theorem}{Theorem}[section]

\newtheorem{corollary}{Corollary}[section]

\newtheorem{lemma}{Lemma}[section]

\newtheorem{proposition}{Proposition}[section]

\newfont{\bbf}{cmbx12 scaled 1435}

\makeatletter
\renewcommand\@biblabel[1]{}
\makeatother
\begin{document}

\begin{titlepage}
\title{Nonparametric Identification of First-Price Auction with Unobserved
Competition: A Density Discontinuity Framework\\
}
\author{Emmanuel Guerre \\
School of Economics and Finance\\
Queen Mary, University of London\\
United Kingdom \and Yao Luo \\
Department of Economics\\
University of Toronto\\
Canada}
\date{May 2026}
\maketitle
\vfill
{\footnotesize  \textbf{Acknowledgments}: We are also grateful to Lu Han, Xavier d'Haultf\oe uille, Elia Lapenta, Daniel Lopes Ribeiro, Quang Vuong, Yuting Wang together with conference and seminar participants for useful discussions and suggestions. Cheok In Fok and Jiaqi Zou provided excellent research assistance. Luo thanks SSHRC for financial support. All  errors are our own.}
\thispagestyle{empty}
\newpage
\thispagestyle{empty}
\begin{abstract}

We consider nonparametric identification of independent private value first-price auction models, in which the analyst only observes winning bids. Our benchmark model assumes an exogenous number of bidders $N$. We show that, if the bidders observe $N$, the resulting discontinuities in the winning bid density can be used to identify the distribution of $N$. The private value distribution can be nonparametrically identified in a second step.  
This extends, under testable identification conditions, to the case where $N$ is a number of potential buyers, who bid with some unknown probability. Identification also holds in  presence of additive unobserved heterogeneity drawn from some parametric distributions. 



\vfill

\noindent

\end{abstract}
\end{titlepage}

\pagebreak


\bigskip

\section{Introduction}


\paragraph{Motivations.} 
There exists a large literature on nonparametric identification of auction models; see, e.g., Athey and Haile (2007) or Hendricks and Porter (2007) for a review. In the case of sealed-bid first-price auctions, a vast majority of work assumes that the analyst can observe all of the bids, or both the winning bid and the number of competitors.
This may not always be observed. In French timber auctions, for example, only the winning bid may be available to researchers to preserve bidder anonymity (Lamy, 2012).
Indeed, it is common practice in many markets that can be treated as auctions for only the winning bid (i.e. the transaction price) to be recorded. For instance, a company soliciting price quotes for a task to be completed is implicitly organizing a first-price auction. While the company may not record all quotes or the number of responses, the price paid to the winning bidder is likely to appear in accounting records. ``Bidding wars'' are becoming commonplace in housing markets, where houses are sold through a competitive bidding process resembling an informal first-price auction, as noted in Han and Strange (2015). Governments may offer subsidies to attract firms, as recently considered by Kim and Yu (2024) and Slattery (2020) using an auction framework. Observing all the subsidy offers may be difficult, because states or firms may both have some interest in confidentiality. Hence, in many economic situations of interest, the records may only contain the final winning bid. Therefore, the ability to identify auction primitives solely from winning bid data may enlarge the scope of auction theory applications.

A second motivation stems from misspecification considerations. Indeed, structural estimation procedures of auctions crucially depend on the number of active buyers, which may differ from the observed number. For instance, Laffont, Ossard and Vuong (1995) consider an application where  bidders are agents of several retail sellers, in which case the number of bids underestimates the level of competition. Given 11 observed bids, these authors estimate the number of active buyers to be 18, causing important changes in the estimated structural parameters. Alternatively, some buyers may enter an auction simply to gain information, in which case their bids would be dominated and never impact the winning bid. Some bidders may collude and place phantom bids in an attempt to hide their cartel memberships;
counting these bids would be misleading as it does not account for collusion.
For instance, in many procurement settings,  sellers can be legally constrained to stop auctions with less than three bidders in attendance. In this case, two buyers may be tempted to contact a third to submit a cover bid but ultimately allow the auction to go through. Econometric methods that check the effectiveness of bidder attendance can be a useful tool for regulators before proceeding to further possibly costly investigations.
More generally, using only winning bids provides a robust approach for identifying primitives of interest when the truthfulness of observed  bids is dubious.

Last, participation is a parameter of interest in itself. 
As noted by Bulow and Klemperer (1996), increasing competitive participation would yield higher seller expected revenues than choosing an optimal reserve price under the symmetric independent private value paradigm. 
In our setup, the number of active buyers is viewed as a latent random variable  which can vary across auctions. Estimating its distribution and comparing it with the observed number of bids when available may be useful to detect participation anomalies.

\paragraph{Baseline model.} Under the independent private value paradigm, buyers bid below their private values. The mechanism of the Bayesian Nash equilibrium implies in turn that the bid probability density function (pdf hereafter) is positive throughout its support, illustrating the attractiveness of such profitable bids. As a consequence, the bid pdf is positive at its upper bound and zero thereafter, resulting in a discontinuity at this point.

We  develop a new approach for identification of first-price auction models that exploits associated discontinuities in the winning bid pdf. First, we identify the distribution of unknown competition. In particular, we build on an important restriction that first-price auction models impose on the data: the bid quantile function must be strictly increasing with respect to the number of bidders. Therefore, the upper boundary of the bid distribution, conditioning on the number of bidders, is strictly increasing, as well. 
As bid densities are discontinuous at these locations, this creates jumps at these upper boundaries in the winning bid pdf.  A novel result of the paper is that these jumps identify the distribution of the number of bidders.

Second, we identify the value distribution function by iteratively exploiting two equilibrium mappings  that relate the value and bid quantile functions. Based on the location of the discontinuities in the winning bid density function, we create a sequence of expanding quantile intervals over which the private value quantile is identified. For every iteration, we start by identifying the bid quantile function in the most competitive auction, which has the largest number of bidders. This information can then be used to identify the value quantile function in the same quantile interval and further calculate the corresponding bid quantile for other competition levels.

\paragraph{Buyer uncertainty and unobserved heterogeneity.}
The paper considers several extensions of the baseline model. Section \ref{BUAH} focuses on buyer uncertainty and auction heterogeneity. As the econometrician in the baseline model, buyers may also face unknown competition, as considered in Harstad, Kagel and Levin (1990) and Kong (2020), among others. Such uncertainty may arise due to the presence of a reserve price, as considered in Guerre, Perrigne and Vuong (2000), or entry costs, as in Li and Zheng (2009) or Marmer, Shneyerov and Xu (2013). We consider a setup where the number of potential bidders, $N$, is known by the buyers but not the effective one. We allow $N$ to vary across auctions, and give conditions on its support ensuring identification of its distribution together with the private value one and the probability that a potential buyer bids.

First-price auctions with unobserved heterogeneity affecting the auctioned good are considered in Krasnokutskaya (2011), and are especially challenging in our framework. Indeed, the presence of a continuous unobserved heterogeneity component washes discontinuities out of the winning bid density. Fortunately, considering its first and second derivatives allows for the identification of the participation distribution. We also show that some features of the unobserved component distribution can be recovered, so that it can be parametrically identified, raising hope for nonparametric identification of the private value distribution.

\paragraph{Related literature.} Allowing for unknown competition started early in the empirical auction
literature. Laffont et al. (1995) estimate the number of buyers $N$ as a
parameter that they take to be constant across auctions. Paarsch (1997) treats unknown competition as a nuisance parameter, which is eliminated using
conditional likelihood estimation. For ascending eBay auctions, Song (2004)
shows that the private value distribution and a constant number of buyers
are identified from winning and second-highest bids, but not from
winning bids alone when $N$ is random. More pertinent to our paper is the misclassification
approach of An, Hu and Shum (2010), who study identification from the winning bid using a
proxy $N^* \leq N$ for the number of buyers and an instrument that can be a
discretized second bid. Shneyerov and Wong (2011) suppose that only
winning bids and the number of active bidders are observed. Some recent work on ascending auctions with unobserved heterogeneity uses the winning bid plus additional information for identification when the number of bidders is imperfectly observed or unobserved. More specifically, Freyberger and Larsen (2022) and Luo and Xiao (2023) use additional bids, while Hern\'andez, Quint and Turansick (2020) assumes that a ``participation shifter'' is available.

The present paper contributes to the literature on nonparametric
identification of finite mixtures; see for instance the review of Compiani
and Kitamura (2016). Existing identification results require either
exclusion restrictions or multiple independent measurements. A first-price auction example of the latter can be found in D'Haultf\oe uille and F\'{e}vrier (2015), who recover the distribution of an unobserved continuous auction characteristic from three bids. Our approach only uses the winning bid.

The discontinuity design (DD) literature has expanded rapidly in recent years; interested
readers are encouraged to refer to review papers by Imbens and Lemieux (2008), Kleven
(2016) and Jales and Yu (2017). Recent auction applications include Coviello
and Marinello (2014), Choi, Neisheim and Rasul (2016) and Kawai, Nakabayashi, Ortner and Chassang (2023). As in the DD
literature, this paper employs jump sizes for identification purposes --- more
specifically, to identify the participation distribution. However the structural nature of our paper departs from this literature.


The remainder of the paper is organized as follows. Section 2 presents the identification results for the baseline model. Section 3 extends the analysis to account for buyer uncertainty and unobserved heterogeneity. Section 4 concludes. Appendix A collects the remaining proofs.

\section{The benchmark model \label{Benchmark}}

In this section, we start by describing the benchmark auction model and introduce two equilibrium mappings that are convenient for describing our discontinuity identification strategy. Next, we derive the restrictions that the model imposes on the observed winning bids, especially with respect to the formation of discontinuities. Finally, we describe our identification strategy in two steps. First, we identify the distribution of the number of buyers from the discontinuities in the winning bid density function. Second, we identify the value distribution function using the two equilibrium mappings iteratively. 

\subsection{The symmetric independent private values paradigm}

Suppose there is a single item for sale with $N$ active symmetric buyers bidding for the
item. All buyers observe $N$. In contrast, the analyst does not observe $N$, which causes auction-specific unobserved heterogeneity. Each buyer $i$ also observes her private valuation $V_{i}$, which is unknown to other buyers. The private values $V_{i}$ are $i.i.d.$ draws from a distribution $F\left(\cdot\right) $, which is known to all the buyers and is independent of $N$. The buyers are risk neutral and their bids $B_{i}$ are formed according to a
symmetric best-response strategy. 
In sum, the primitives are the distribution of the number of buyers $N$ and the private value distribution.

We assume that the analyst only observes the winning bid $W$, i.e., the maximum
bid among the $N$ buyers in the set $\mathcal{N}$ of active buyers 
\begin{equation*}
W =\max_{i \in \mathcal{N}}B_{i}.
\end{equation*}
Hence, the analyst observes draws from the unconditional cumulative probability
distribution of the winning bid $G(\cdot)$,
which is a mixture of the conditional winning bid distributions given $N$:
\begin{eqnarray}
G(b) & = & \sum_{n=2}^{+\infty} \mathbb{P} \left( \max_{1 \leq i \leq n
} B_i \leq b \right) \times \mathbb{P} \left( N=n \right) = \sum_{n=2}^{+\infty} G_{n}^{n}\left( b\right) \mathbb{P} \left( N=n \right), 
\label{Mixture}
\end{eqnarray}
where $G_n(\cdot)$ is the conditional bid distribution given $N=n$.

The following two assumptions introduce some additional conditions for the
distribution of $N$ and for the private value distribution $F\left(
\cdot\right) $.

\bigskip

\textbf{Assumption N}.\textit{\ The number of active buyers }$N$%
\textit{\ is a discrete random variable with support }$\left\{ \underline{n}%
,\ldots,\overline{n}\right\} $ \textit{for some integers }$2\leq%
\underline{n}\leq\overline{n}<\infty$\textit{, i.e., }$p_{n}=\mathbb{P}\left(
N=n\right) >0$ \textit{for }$n=\underline{n},\underline{n}+1,\ldots,%
\overline{n}$ \textit{with }$\sum_{n=\underline{n}}^{\overline{n}}p_{n}=1$.

\bigskip

\textbf{Assumption IPV}. \textit{Buyers' private values $V_{i}$ are $i.i.d.$ draws from a common knowledge distribution $F\left(\cdot\right)$ and are unknown to competitors. The cumulative distribution function $F\left(\cdot\right)$ has a
compact support }$\left[ \underline{v},\overline{v}\right] $. \textit{Its
probability density function }$f\left( \cdot\right) $ \textit{is continuous
and strictly positive over }$\left[ \underline{v},\overline {v}\right] $. 


\bigskip

Both the theoretical and empirical literature adopt the assumption of a private value distribution with compact support. In particular, it 
rules out multiple asymmetric equilibria; see Maskin and Riley (1984, Remark
2.3), who also establish that symmetric Bayesian Nash Equilibrium bids are
given by a strictly increasing and continuously differentiable function of private values. 

For our discontinuity approach, the compact support assumption ensures the existence of discontinuities in the density of unconditional winning bids that we exploit in this paper. In particular, the winning bid densities $g_n (\cdot)$ given $N=n$ stay bounded away from $0$ at the upper boundary of its support; see (\ref{Uppergn}) below.

That the private value pdf $f(\cdot)$ is positive over $[\underline{v},\overline{v}]$ is a current assumption in the auction literature, as seen in Maskin and Riley (1984), Lebrun (1999), Guerre et al. (2000), among others. It is used here to identify the lowest number $\underline{n}$ of active buyers; see Lemma \ref{Idprelims}-(iii) below. The alternative identification method of Section \ref{BU} allows us to relax this condition as noted in Footnote \ref{Notail}.

\subsection{Bid and value quantile equilibrium mappings }

In this subsection, we describe two equilibrium mappings that are repeatedly used in our identification procedure. Specifically, there is an equilibrium mapping from the value distribution to the bid distribution, and vice versa. Our discontinuity identification strategy is conveniently described using the quantile framework as in Guerre, Perrigne and Vuong (2009), Liu and Luo (2017), and Guerre and Gimenes (2022), that we recount below.

Let $V\left(
\alpha\right)=F^{-1}\left( \alpha\right) $ represent the private value quantile
function, where $\alpha \in \left[ 0,1\right] $ is the quantile level. Let $%
B_{n}\left( \alpha\right) $ denote the bid quantile function given that $n$ buyers
participate in the auction, and set $B_n^{(1)} (\alpha) = \frac{\partial B_n (\alpha)}{\partial \alpha}$. Following Milgrom (2001)'s exposition of the
identification strategy of Guerre, Perrigne and Vuong (2000), the private
value quantile function $V\left( \cdot\right) $ can be viewed as the common
valuation function of buyers who receive independent uniform private signals 
\begin{equation*}
A_{i}=F\left( V_{i}\right), 
\end{equation*}
which determines their private values $V_{i}=V\left( A_{i}\right) $. By Assumption
IPV, $B_{i}=\beta_{n}\left( A_{i}\right) $ for all $i$, where $
\beta_{n}\left( \cdot \right)$ is strictly increasing and continuously
differentiable. It follows that for any $b$ in the range of $\beta_{n}\left(
\cdot\right) $,%
\begin{equation*}
G_{n}\left( b\right) =\mathbb{P}\left( \beta_{n}\left( A_{i}\right) \leq
b\right) =\mathbb{P}\left( A_{i}\leq\beta_{n}^{-1}\left( b\right) \right)
=\beta_{n}^{-1}\left( b\right),
\end{equation*}
because $A_{i}$ is uniformly distributed over $\left[ 0,1\right] $. 
Hence, the best-response strategy is the bid quantile function
\begin{equation*}
\beta_{n}\left( \alpha\right) =B_{n}\left( \alpha\right) \text{ for all }%
\alpha \in \left[ 0,1\right].
\end{equation*}

Now, let us relate the bid and private value quantile functions.
Suppose that buyer $i$ receives signal $\alpha$ but makes a generic
bid $B_{n}\left( a\right) $ for some $a \in \left[ 0,1\right] $. Since her
opponents bid $B_{n}\left( A_{j}\right) $, the probability that her bid $%
B_{n}\left( a\right) $ wins the auction is given by $\mathbb{P}(\max_{j\neq
i}A_{j}\leq a)$, which is equal to $a^{n-1}$ as the signals of the $%
n-1$ opponents $A_{j}$, where $j \neq i$, are independent and uniform. It follows that the expected
payoff of buyer $i$ is $\left( V\left( \alpha\right) -B_{n}\left( a\right)
\right) a^{n-1}$, which is maximized when $a=\alpha$. Since
\begin{align*}
\left. \frac{\partial}{\partial a}\left[ \left( V\left( \alpha\right)
-B_{n}\left( a\right) \right) a^{n-1}\right] \right\vert _{a=\alpha} &
=
V\left( \alpha\right)
\left(
n-1\right) \alpha^{n-2}
-
\frac{\partial 
\left[
B_{n}\left( \alpha\right) \alpha^{n-1}
\right]
}{\partial \alpha}
\\
& =\left( n-1\right) \alpha^{n-2}\left( V\left( \alpha\right) -B_{n}\left(
\alpha\right) -\frac{\alpha B_{n}^{\left( 1\right) }\left( \alpha\right) }{%
n-1}\right), 
\end{align*}
setting this derivative to $0$ gives
\footnote{It gives equivalently that $B_n^{(1)} (\alpha)=\left(V(\alpha) -B_n (\alpha)\right)\cdot \frac{n-1}{\alpha} <\infty$. As $g_n(b)=1/B_n^{(1)}\left[G_n (b)\right]$, it follows that a Bayesian Nash equilibrium bid pdf $g_n(\cdot)$ cannot vanish on $(\underline{v},\overline{b}_n]$, as mentioned in the introduction.}
\begin{equation}
V\left( \alpha\right)  = B_{n}\left( \alpha\right) +\frac{\alpha B_{n}^{\left( 1\right) }\left(
\alpha\right) }{n-1} .  \label{Nash}
\end{equation}
This constitutes the equilibrium mapping from the bid quantile function to the value quantile function, which is the basis of the identification of $V(\cdot)$ with knowledge of $B_n(\cdot)$. 

Now, let us consider the inverse of the mapping (\ref{Nash}). Indeed, (\ref{Nash}) is equivalent to
$\frac{\partial 
\left[
B_{n}\left( \alpha\right) \alpha^{n-1}
\right]
}{\partial \alpha}=V\left( \alpha\right)
\left(
n-1\right) \alpha^{n-2}$, and it follows 
\begin{equation}
B_{n}\left( \alpha\right)
=
\frac{n-1}{\alpha^{n-1}}
\int_0^{\alpha}
t^{n-2} V(t) dt.
\label{Bqf}
\end{equation}
For convenience of identification that will be clarified later on, let us introduce the conditional  bid upper bound 
\[
\overline{b}_n 
=
B_n (1)
=  
(n-1)
\int_0^{1}
t^{n-2} V(t) dt , 
\] 
which gives 
\begin{equation} \label{Nash2}
B_n(\alpha) = \frac{n-1}{\alpha^{n-1}}\left[ \frac{\overline{b}_{n}}{n-1}-\int_{\alpha}^{1}t^{n-2}V\left( t\right) dt\right].
\end{equation}
This constitutes the equilibrium mapping from the value quantile to the bid quantile function. 
The two mappings represented in (\ref{Nash}) and (\ref{Nash2}) are repeatedly used in our identification procedure.  

\subsection{Structure of the winning bid distribution}

The structure of winning bid distributions compatible with a first-price auction where buyers observe $N$ follows from the mixture expression of $G(\cdot)$ in Equation (\ref{Mixture}) and the best-response differential equation (\ref{Nash}).

\begin{proposition}
\label{GIPVNknown} A cdf $G\left( \cdot\right) $ is rationalized by a
first-price auction model satisfying Assumptions N, IPV, and observability of $N$ by buyers if and only if

\begin{enumerate}
\item The cdf $G\left( \cdot\right) $ has a mixture structure 
\begin{equation}
G\left( \cdot\right) = \sum_{n=\underline{n}}^{\overline{n}}p_{n}G_{n}
^{n}\left( \cdot\right) , \label{GNknown}
\end{equation}
where the $G_n(\cdot)$ are cdf, $2 \leq \underline{n} \leq \overline{n}$,
and the positive $p_n$ satisfy $\sum_{n=\underline{n}}^{\overline{n}}p_{n}=1$%
.

\item The quantile functions $B_{n}\left( \cdot\right) =G_{n}^{-1}\left(
\cdot\right) $ are continuously differentiable over $[0,1]$ and satisfy the
compatibility conditions 
\begin{equation*}
B_{n} \left( \alpha\right) +\frac{\alpha B_{n}^{\left( 1\right) }\left(
\alpha\right) }{n-1}=B_{m}\left( \alpha\right) +\frac{\alpha B_{m}^{\left(
1\right) }\left( \alpha\right) }{m-1} 
\end{equation*}
for all $\underline{n} \leq n,m \leq \overline{n}$ and all $\alpha \in [0,1] $. Moreover, the function $V(\alpha)=B_{n} \left( \alpha\right) +\frac{\alpha B_{n}^{\left( 1\right) }\left(
\alpha\right) }{n-1}$ is continuously differentiable
over $[0,1]$ with $V^{(1)} (\cdot)>0$.
\end{enumerate}
\end{proposition}

In short, a cdf $G(\cdot)$ as in Proposition \ref{GIPVNknown} is a
mixture with components constrained by compatibility conditions driven by
the best response differential equation (\ref{Nash}).
The compatibility conditions of Proposition \ref{GIPVNknown}-(ii) reflect that the mixture components $G_n (\cdot)$ are generated by the same private value distribution, an important feature for identification. In particular, our identification results rely on the constraints it imposes on the extremities of the conditional bid pdf $g_n (\cdot)$, as illustrated in the next corollary. 
Recall
$\overline{b}_n = B_n (1)$, $\underline{v}=V(0)=\underline{b}_n$, and $\overline{v}=V(1)$.
\begin{corollary}
Suppose that the compatibility conditions of Proposition \ref{GIPVNknown}-(ii) hold.  Then, for all $n=\underline{n},\ldots, \overline{n}$, we have $\overline{b}_{n}<\overline {v}$ together with:
\begin{align}
g_{n}\left( \underline{v}\right) & =\frac{n}{n-1}f\left( \underline{v}%
\right) , \mbox{ with }  \label{Lowergn} \\
g_{n}\left( \overline{b}_{n}\right) & =\frac{1}{(n-1)(\overline {v}-%
\overline{b}_{n})}.  \label{Uppergn}
\end{align}
\label{Extgn}
\end{corollary}

Equation (\ref{Uppergn}) implies that  $g_{n}\left( \overline{b}_{n}\right)$ is strictly positive. It turns out from (\ref{Mixture}) that this causes discontinuities in the winning bid pdf $g(\cdot)$ at each  $\overline{b}_{n}$, as studied in the next section. As it follows that $\overline{b}_{n}$ is identified, (\ref{Uppergn}) shows that $g_{n}\left( \overline{b}_{n}\right)$, where $n\in \{\underline{n},\ldots,\overline{n}\}$, are determined by the common unknown parameter $\overline{v}$. We employ this consequence of the compatibility conditions of Proposition \ref{GIPVNknown}-(ii) later on to identify the distribution of $N$.

\subsection{Winning bid density discontinuities}

In this subsection, we introduce a numerical example to illustrate the discontinuity features of the winning bid pdf that follows from Corollary \ref{Extgn}. This example will also be useful for introducing our identification procedure. A general lemma completes the example.

\subsubsection{Numerical example\label{Example}}
Consider the private value cdf $F(v) = v^2$ for all $v$ in $[0,1]$ and a number of buyers $N=\{2,3\}$ with equal probability. As $V(\alpha)=\alpha^{1/2}$, it follows that 
\[
B_n (\alpha)= \frac{n-1}{\alpha^{n-1}} \int_0^{\alpha} t^{n-2+\frac{1}{2}}dt
=
\frac{n-1}{n-\frac{1}{2}}\alpha^{1/2}.
\]
Hence, $\overline{b}_n=\frac{n-1}{n-\frac{1}{2}}$ yields the conditional bid pdf. $g_n (b)$, given $N=n$, is equal to  $2b/\overline{b}_n^2$ on $[0,\overline{b}_n]$ and vanishes outside this interval.  
Note that the support of the conditional density function increases with the number of buyers. As expected from (\ref{Uppergn}), the pdf $g_2 (\cdot)$ and $g_3 (\cdot)$ are positive below their respective upper boundaries
$\overline{b}_2$ and $\overline{b}_3$, and vanish beyond.

Let us now turn to the winning bid, the observation of the analyst. As expected, the unconditional pdf
$
g(b) = \frac{1}{2} \cdot  2 G_2(b)g_2(b) + \frac{1}{2} \cdot 3 G_3^2(b)g_3(b) 
$
displayed in Figure \ref{Pdf}
is discontinuous at $\overline{b}_2$ and $\overline{b}_3$, with jump sizes $\Delta_2$ and $\Delta_3$, respectively.
The resulting winning bid cdf exhibits kinks at these values, as illustrated in Figure \ref{Cdf}. In this example, Figure \ref{Pdf} exhibits two discontinuities (and Figure \ref{Cdf} exhibits two kinks) because $N$ takes two potential values here. 

\begin{figure}[h]
\begin{subfigure}[b]{0.3\textwidth}
\begin{tikzpicture}[xscale=5,yscale=2.5]
\draw [ultra thick, <->] (0,1.5) -- (0,0) -- (1.1,0);
\node [below right] at (1.1,0) {$b$};
\node [above left] at (0,1.2) {$G (b)$};
\draw[dashed, very thick] (0,1) -- (4/5,1);
\node[left] at (0,1) {$1$};
\draw[blue, ultra thick, domain=0:(2/3)] plot (\x,{pow(3*\x/2,4)/2+pow(5*\x/4,6)/2});
\draw[blue, ultra thick, domain=(2/3):(4/5)] plot (\x,{pow(5*\x/4,6)/2+1/2});
\draw[blue, ultra thick, domain=(4/5):(1.0)] plot (\x,{1});
\node [below, blue] at (0,0) {$\underline{v}$};
\node [below, blue] at (2/3,0) {$\overline{b}_2$};
\draw [dashed, very thick, blue]
(0,0.6674) -- (2/3,0.6674);
\node [left, blue] at (0,0.6674) {$\frac{1}{2}(1+G_3^3(\overline{b}_2))$};
\node [below, blue] at (4/5,0) {$\overline{b}_3$};
\draw[dashed, very thick, blue] (2/3,0) -- (2/3,0.6674);
\draw[dashed, very thick, blue] (4/5,0) -- (4/5,1);
\end{tikzpicture}
\caption{cdf $G(\cdot)$}
\label{Cdf}
\end{subfigure}
\quad\quad\quad \quad\quad\quad \quad\quad\quad 
\begin{subfigure}[b]{0.3\textwidth}
\begin{tikzpicture}[xscale=5,yscale=0.9]
\draw [ultra thick, <->] (0,5) -- (0,0) -- (0.9,0);
\node [below right] at (0.9,0) {$b$};
\node [above left] at (0,5) {$g (b)$};
\draw[blue, ultra thick, domain=0:(2/3)] plot (\x,{10.125*pow(\x,3)+11.44*pow(\x,5)});
\draw[blue, ultra thick, domain=(2/3):(4/5)] plot (\x,{11.44*pow(\x,5)});
\node [below, blue] at (0,0) {$\underline{v}$};
\node [below, blue] at (2/3,0) {$\overline{b}_2$};
\node [below, blue] at (4/5,0) {$\overline{b}_3$};
\draw[dashed, very thick, blue] (2/3,0) -- (2/3,4.51);
\draw[dashed, very thick, blue] ((2/3,4.51) -- (0,4.51);
\draw [decorate,decoration={brace,amplitude=10pt},xshift=0pt,yshift=0pt]
(0,1.51) -- (0,4.51) node [blue,midway,xshift=-0.6cm] 
{ \textcolor{blue}{$\Delta_2$}};
\draw[dashed, very thick, blue] (4/5,0) -- (4/5,15/4);
\draw [decorate,decoration={ brace,amplitude=10pt,mirror},xshift=0pt,yshift=0pt]
(4/5,0) -- (4/5,15/4) node [blue,midway,xshift=0.6cm] 
{ \textcolor{blue}{$\Delta_3$}};
\draw[dashed, very thick, blue] (0,1.51) -- (2/3,1.51);
\end{tikzpicture}
\caption{pdf $g(\cdot)$}
\label{Pdf}
\end{subfigure}
\caption{Winning bid distribution ($V(\protect\alpha)=\protect\sqrt{\protect%
\alpha}$ and $\mathbb{P}(N=2)=\mathbb{P}(N=3)=1/2$)}
\label{Distw}
\end{figure}

\subsubsection{The general case}

The increasing support property of the conditional bid pdf and the winning bid pdf discontinuities in Figure \ref{Pdf} are generic, as shown in the upcoming lemma.
Lemma \ref{Idprelims}-(i) states more generally that bids increase with competition --- a key feature of first-price auctions that does not hold in ascending ones, or when buyers do not observe $N$. Lemma \ref{Idprelims}-(ii) focuses on the winning bid pdf discontinuities and its jumps. 
The identification of $\underline{n}$ in Lemma \ref{Idprelims}-(iii) uses that the lower tail  index of $G(b)$ is the one of $\left(G_{\underline{n}} (b)\right)^{\underline{n}}$, which is $\underline{n}$. 

\begin{lemma}
\label{Idprelims} Suppose Assumptions N and IPV hold. Then, all of the following hold. 
\begin{enumerate}
\item For all $\alpha$ in $\left( 0,1\right] $, $B_{\underline{n}}\left(
\alpha\right) <\cdots< B_{\overline {n}}\left( \alpha\right) < V(\alpha) $
with $B_{n}\left( 0\right) =V\left( 0\right) $ for all $n$. In particular,
for $\overline{b}_n = B_n (1)$, $\overline{b}_{\underline{n}}<\cdots <%
\overline{b}_{\overline{n}}<\overline{v} $.

\item The cdf $G\left( \cdot\right) $ has a pdf $g\left( \cdot\right) $
with $g(\underline{v})=0$, which is continuous on the real line with
the exception of the $\overline{n}-\underline{n}+1$ discontinuity points $%
\overline {b}_{\underline{n}}<\cdots<\overline{b}_{\overline{n}}$, with the
interval $\left[ \underline{v},\overline{b}_{\overline{n}}\right] $ being
the support of $G\left( \cdot\right) $. For $\underline{n} \leq n \leq 
\overline{n}$, the jumps $\Delta_n =\lim_{t\downarrow 0}\left( g(\overline{b}_n-t)-g(\overline{b}
_n+t) \right)$ satisfy 
\begin{equation}  \label{Jumps}
\Delta_n = \frac{np_n}{(n-1)(\overline{v}-\overline{b}_{n})}.
\end{equation}

\item It holds that $\underline{n}=\lim_{t\downarrow0}\frac{\log G\left( 
\underline{v}+t\right) }{\log t}$.
\end{enumerate}
\end{lemma}

Lemma \ref{Idprelims} is an important building block for identifying the competition distribution. Part (iii) is a tail identification result for $\underline{n}$ similar to Hill and Shneyerov (2013) who considered a common value framework.
Lemma \ref{Idprelims}-(ii) shows that the jumps in the winning bid pdf identify $\mathbb{P}(N=n)$ up to the unknown $\overline{v}$.

\subsection{Identification}

Here, we first focus on identification of the participation distribution and then turn to the private values. 

\subsubsection{Identification of the distribution of $N$}

In this subsection, we describe the identification of the support of $N$ and its distribution using the discontinuity points and jump sizes. 
To identify the support, we exploit two implications of Lemma \ref{Idprelims}: (a) the minimum number of buyers $\underline{n}$ is identified from the winning bid distribution tail near the lower boundary; (b) each number of buyers $n$ generates a discontinuity in the winning bid distribution, which identifies the difference $\overline{n}-\underline{n}$.
More specifically, Lemma \ref{Idprelims}-(ii) identifies $\underline{n}$ and $\overline{n}$ through $\underline{n}=\lim_{t\downarrow0}\frac{\log G\left( 
	\underline{v}+t\right) }{\log t}$ and
\[
\overline{n}=\underline{n} +\mathsf{Card}\left\{ b;g\left( \cdot\right) 
\text{ is discontinuous at }b\right\} -1. 
\]
This also identifies the support of the distribution of $N$ as $\mathbb{P}(N=n)>0$ for all $n$ with $\underline{n} \leq n \leq \overline{n}$ by Assumption N.

Next, we exploit the jumps in the pdf to identify $p_n=\mathbb{P} (N=n)$.
Recall that Equation (\ref{Jumps}) identifies $p_n$ up to the private value upper bound  $\overline{v}$, 
\begin{equation*}
	p_{n}=\frac{n-1}{n}\Delta_{n}\left( \overline{v}-\overline{b}_{n}\right) .
\end{equation*}
But $\sum_{\underline{n}}^{\overline{n}}p_{n}=1$ implies  
\begin{equation}
	\overline{v}=\frac{1+\sum_{n=\underline{n}}^{\overline{n}}\frac{n-1}{n}%
		\Delta_{n}\overline{b}_{n}}{\sum_{\underline{n}}^{n=\overline{n}}\frac{n-1}{n}%
		\Delta_{n}} .
	\label{Identbarv} 
\end{equation}
Hence, $p_{n}$ satisfies 
\begin{equation}
	p_{n}
	=
	\frac{
		\frac{n-1}{n}\Delta_{n}
	}{
		\sum_{k=\underline{n}}^{\overline{n}}
		\frac{k-1}{k}\Delta_{k}}
	+
	\frac{n-1}{n}\Delta_{n}
	\left( 
	\frac{
		\sum_{k=\underline{n}}^{\overline{n}}
		\frac{k-1}{k}\Delta_{k}\overline{b}_{k}}{
		\sum_{k=\underline{n}}^{\overline{n}}
		\frac{k-1}{k}\Delta_{k}}
	-\overline{b}_{n}\right) ,\quad 
	n=\underline{n}%
	,\ldots,\overline{n},
	\label{Identpn}
\end{equation}
and is identified because the discontinuity points $\overline{b}_k$ and jump sizes $\Delta_k$ are identified. We summarize these identification results in the next lemma.

\begin{lemma}
	\label{Discontinuity} Suppose Assumptions N and IPV hold. Then $\underline{v}
	$, $\overline{n}$, $\underline{n}$, $\overline{b}_{\underline{n}}<\cdots<%
	\overline{b}_{\overline{n}}$, $\overline{v}$, and the probabilities $p_n$, $n=%
	\underline{n},\ldots,\overline{n}$, are identified.
\end{lemma}

The identifying equations (\ref{Identbarv}) and (\ref{Identpn}) can also be used to derive inequality constraints satisfied by the jumps sizes $\Delta_n$, discontinuity locations $\overline{b}_n$, and the lowest and largest numbers of bidders $\underline{n}$ and $\overline{n}$. Indeed $\overline{v}>\overline{b}_{\overline{n}}$ and $0 \leq p_n \leq 1$ are equivalent to the following inequalities
\begin{align*}
	&
	\sum_{k=\underline{n}}^{\overline{n}}
	\frac{k-1}{k} \Delta_k
	\left(
	\overline{b}_{\overline{n}}-\overline{b}_k
	\right)
	\leq 
	1,
	\\
	&
	1 +
	\sum_{k=\underline{n}}^{\overline{n}}
	\frac{k-1}{k} \Delta_k
	\left(
	\overline{b}_{k}-\overline{b}_n
	\right)
	\leq
	\frac{
		\sum_{k=\underline{n}}^{\overline{n}}
		\frac{k-1}{k} \Delta_k
	}{
		\frac{n-1}{n} \Delta_n
	},
	\quad
	n = \underline{n}, \ldots, \overline{n}, 
\end{align*}
given that $\Delta_n>0$ must also hold by Lemma \ref{Idprelims}-(ii). A violation of any of these inequalities indicate that the model is misspecified.

\subsubsection{Identification of the private value quantile function \label%
		{Ipvqf}}
	
We first return to the numerical example to illustrate our iterative identification procedure for the private value distribution.

\medskip

\paragraph{Numerical example (cont'd).}
By Lemma \ref{Discontinuity}, $\underline{n}=2$, $\overline{n}=3$ and $p_1=p_2=\frac{1}{2}$ are identified.
Let us now turn to the identification of the private value distribution, which is based on the winning bid cdf
\[G(b)=\frac{1}{2}\left( G_2^2(b)+G_3^3(b)\right)\] displayed in Figure \ref{Cdf}. Since $G_2^2(b)=1$ on $[\overline{b}_2,\overline{b}_3]$,
\[
G_3 (b) = \left(2G(b)-1\right)^{\frac{1}{3}},
\quad
b \in  [\overline{b}_2,\overline{b}_3].
\]
It follows that $B_3 (\cdot)$ is identified on $[\alpha_1,1]$, where $\alpha_1=G_3 (\overline{b}_2)$, using the top portion of the winning bid distribution; see Figure \ref{Cdf} when $G(b) \in [\frac{1}{2}(1+G_3^3(\overline{b}_2)),1]$. Using the mapping (\ref{Nash}) from the bid quantile function to the private value one gives 
\[
V(\alpha) = B_3 (\alpha) + \frac{1}{2}\alpha B_3^{(1)} (\alpha), 
\] and $V(\cdot)$ is  identified on $[\alpha_1,1]$. As private values do not depend upon $N$, using the mapping (\ref{Nash2}) from the private value quantile function gives
\[
B_2 (\alpha)
=
\frac{1}{\alpha}
\left[
\overline{b}_2
-
\int_{\alpha}^{1} V(t) dt
\right]
\]
so that $B_2(\cdot)$ is  also identified on $[\alpha_1,1]$. The identified $B_2 (\alpha)$, $B_3 (\alpha)$, and $V (\alpha
)$, where $\alpha \in [\alpha_1,1]$, are displayed in blue in Figure \ref{IdBV}.

\begin{figure}[ht]
	\centering
	\begin{tikzpicture}[xscale=8,yscale=8]
	\draw [ultra thick, <->] (0,1.1) -- (0,0) -- (1.1,0);
	\node [below right] at (1.1,0) {$\alpha$};
	\node [left] at (0,1.1) {$B_n (\alpha)$};
	\draw[yellow, ultra thick, domain=0:0.233] plot (\x, {pow(\x,1/2)});
	\draw[orange, ultra thick, domain=0.233:0.334] plot (\x, {pow(\x,1/2)});
	\draw[red, ultra thick, domain=0.334:(625/1296)] plot (\x, {pow(\x,1/2)});
	\draw[purple, ultra thick, domain=(625/1296):(25/36)] plot (\x, {pow(\x,1/2)});
	\draw[blue, ultra thick, domain=(25/36):1] plot (\x, {pow(\x,1/2)});
	\node[right] at (1,1) {$V(\alpha)$};
	\draw[yellow, ultra thick, domain=0:0.233] plot (\x, {2*pow(\x,1/2)/3});
	\draw[orange, ultra thick, domain=0.233:0.334] plot (\x, {2*pow(\x,1/2)/3});
	\draw[red, ultra thick, domain=0.334:(625/1296)] plot (\x, {2*pow(\x,1/2)/3});
	\draw[purple, ultra thick, domain=(625/1296):(25/36)] plot (\x, {2*pow(\x,1/2)/3});
	\draw[blue, ultra thick, domain=(25/36):1] plot (\x, {2*pow(\x,1/2)/3});
	\node[right] at (1,2/3) {$n=2$};
	\draw[yellow, ultra thick, domain=0:0.233] plot (\x, {4*pow(\x,1/2)/5});
	\draw[orange, ultra thick, domain=0.233:0.334] plot (\x, {4*pow(\x,1/2)/5});
	\draw[red, ultra thick, domain=0.334:(625/1296)] plot (\x, {4*pow(\x,1/2)/5});
	\draw[purple, ultra thick, domain=(625/1296):(25/36)] plot (\x, {4*pow(\x,1/2)/5});
	\draw[blue, ultra thick, domain=(25/36):1] plot (\x, {4*pow(\x,1/2)/5});
	\node[right] at (1,4/5) {$n=3$};
	\draw[dashed, very thick] (1,1) -- (1,0);
	\draw[dashed, very thick, blue] (0,4/5) -- (1,4/5);
	\node[left, blue ] at (0,4/5) {$\overline{b}_3$};
	\node[below] at (0,0) {$0$};
	\node[below] at (1,0) {$1$};
	\draw[dashed, very thick, blue] (0,2/3) -- (1,2/3);
	\node[left, blue] at (0,2/3) {$\overline{b}_2$};
	\draw[dashed, very thick, blue] (25/36,2/3) -- (25/36,0);
	\node[below, blue] at (25/36,0) {$\alpha_1=G_3(\overline{b}_2)$};
	\draw[dashed, very thick, blue] (25/36,5/9) -- (0,5/9);
	\node[blue, left] at (0,5/9) {$\beta_1$};
	\draw[dashed, purple, very thick] (625/1296,5/9) -- (625/1296,0);
	\node[purple, below] at (625/1296,0) {$\alpha_2$};
	\draw[dashed, very thick, purple] (625/1296,25/54) -- (0,25/54);
	\node[purple, left] at (0,25/54) {$\beta_2$};
	\node[red, below] at (0.334,0) {$\alpha_3$};
	\draw[dashed, very thick, red] (0.334,25/54) -- (0.334,0);
	\node[red, left] at (0,0.386) {$\beta_3$};
	\draw[dashed, very thick, red] (0.334,0.386) -- (0,0.38);
	\node[orange, below] at (0.233,0) {$\alpha_4$};
	\draw[dashed, very thick, orange] (0.233,0.386) -- (0.233,0);
	\node[orange, left] at (0,0.322) {$\beta_4$};
	\draw[dashed, very thick, orange] (0.233,0.322) -- (0,0.322);
	\end{tikzpicture}
	\caption{Iterative identification of $B_n(\protect\alpha)$ and $V(\protect%
		\alpha)$ from $G(\cdot)$, as in Figure \protect\ref{Distw}.}
	\label{IdBV}
\end{figure}

Next, we enlarge the interval $[\alpha_1,1]$ over which $V(\cdot)$ is identified. For this purpose, let $\beta_1=B_2 (\alpha_1)<\overline{b}_2$ and observe that $G_2(b)$ is identified for $b \geq \beta_1$. Given that 
\[
G_3 (b) = \left(G_2^2(b)-2G(b)\right)^{\frac{1}{3}},
\]
$G_3 (b)$ is identified for $b \geq \beta_1$, as $B_3 (\alpha)$ is identified for $\alpha \geq \alpha_2=G_3 (\beta_1)$. Figure \ref{IdBV} shows that $\alpha_2<\alpha_1$ and arguing as above gives us identification of $V(\cdot)$ and $B_2 (\cdot)$ on $[\alpha_2,1]$. Three portions of $V(\cdot)$, $B_3 (\cdot)$, and $B_2(\cdot)$ are identified through three iterations and plotted in Figure \ref{IdBV} in purple, red, and orange, respectively. Furthermore, Figure \ref{IdBV} suggests that additional iterations of this identification procedure should allow us to recover any $V(\alpha)$.

\paragraph{The general case.} The iterative identification described above can be easily generalized. As in Guerre et al (2009), competition (i.e., that the bid quantile function increases with $n$ and only cross at $0$) ensures that the quantile-level sequence $\{\alpha_k\}$ decreases to $0$, allowing identification of the full private value distribution. This implies identification of the private value distribution when only observing the winning bid, as stated in the next general result.

\begin{theorem}
	\label{FP} Suppose Assumptions N and IPV hold and that the buyers observe
	the number of active buyers $N$. Then $F\left( \cdot\right) $ and the
	distribution of $N$ are identified.
\end{theorem}

\subsection{Auction covariate and estimation strategy}

\paragraph{Auction covariate.}
Many auctioned goods depend upon observed covariate that should also affect the private values. A popular specification from Haile, Hong and Shum (2003), Rezende (2008), considers a regression specification  \[\widetilde{V}_i = X^{\prime} \theta + V_i.\] 
It follows that 
\[
\widetilde{W} = X'\theta + W, \text{ where } W = \max_{i \in \mathcal{N}} B_i,
\]
where the distribution of the bid $B_i$ is identical to the one of the baseline model, provided the private values and the set $\mathcal{N}$, and in particular the number of active buyers $N$, are independent of $X$. Under this assumption, $\theta$ is identified regressing $\widetilde{W}$ on $X$ and an intercept, allowing for the recovery of $W$ and the identification approach of the baseline model to proceed. 

It is however crucial here that the number of potential buyers, $N$, is independent of $X$. Otherwise, the regression above may identify a biased version $\theta^{\star}$ of $\theta$ which captures the possible correlation of $\mathbb{E} \left[W\right]$, which depends upon $N$, and $X$. The corresponding regression error is then $W^{\star} = X^{\prime} \left(\theta-\theta^{\star}\right) + W$. If $X$ has a continuous distribution, so will  $X^{\prime} \left(\theta-\theta^{\star}\right)$ when $\theta \neq \theta^{\star}$. The expression (\ref{Tildeg}) of the density of the winning bid contaminated with continuous unobserved heterogeneity then suggests that the pdf of $W^{\star}$ will have no discontinuities even if $N$ varies, so that the proposed identification approach does not apply.

\paragraph{Estimation strategy.}
In the presence of covariate, a two-step estimation method as above may blur discontinuities as described above, and therefore leads to further inconsistencies. Building upon our identification strategy seems unrealistic: while estimating discontinuity locations and sizes is theoretically feasible (see, among many others, Chu and Cheng (1996), who deal with both when the number of jump points is unknown), the iterative nature of the private value distribution identification renders its implementation difficult. 

One-shot estimation in the fashion of Guerre et al (2000) cannot be applied directly due to the lack of a known number of bidders and a proper bid distribution. An alternative option is to rely on a likelihood estimation approach, such as the Bayesian ones implemented empirically to auctions by  Li and Zheng (2009) and Aryal,  Charankevich, Jeong and Kim (2023) among others. Theoretical studies in Hirano and Porter (2003) or Chernozhukov and Hong (2004) have illustrated efficiency and feasibility of the parametric Bayesian approach for nonregular models, including first-price auctions under the independent private value paradigm with a known number of bidders. In a parametric framework, the latter model induces bids which upper support bounds depend upon the parameter of interest, with a positive pdf jumping to $0$ at this location.

These works do not consider however the case of a pdf exhibiting several jumps as for our winning bid distribution. This case was addressed in Ibragimov and Has'minskii (1981, chapter 5) in the absence of covariate, while Ghosal and Samanta (1995) allowed for covariate. However, the number of discontinuity must be constant across parameters, meaning that in our framework, the difference $\overline{n}-\underline{n}$ of the maximal and minimal number of bidders should be known.

\section{Buyer uncertainty and auction heterogeneity \label{BUAH}} 
\subsection{Buyer uncertainty \label{BU}}

\paragraph{Setup and assumptions.}
Consider an environment where there are $N$ potentially active buyers, who submit a bid with probability $d$, an additional parameter to be identified. Buyer uncertainty then arises from the fact that the total number of active buyers is not known but follows a binomial distribution of parameter $(N,d)$ given $N$, assuming from now on that the bidding decisions and $N$ are independent. This auction setup is summarized in the next assumption.

\bigskip

\textbf{Assumption BU}.\textit{\ There are $N$ potentially active buyers, observable to the buyers but not the econometrician. Given $N$, each buyer  decides to participate in the auction with probability $d$ in $(0,1)$, privately and independently of the other buyers. Each active buyer draws a private value from the common knowledge distribution $F(\cdot)$ and the seller reserve price is $\underline{v}>0$. The econometrician observes the winning bid $W$ or that the auction fails if none of the buyers attend.} 

\bigskip

In addition, the private value distribution $F(\cdot)$ satisfies Assumption IPV, and $N$ can vary across auctions with $p_n = \mathbb{P} (N=n)$, $n=\underline{n},\ldots,\overline{n}$ as in Assumption $N$. 
This setup covers in particular the case where there is a reserve price $r$ constant across auction, $F(\cdot)$ denoting the truncated private value distribution $F(\cdot|V \geq r)$ and $d=P(V\geq r)$ being the probability of entering the auction.

\paragraph{Model primitives.}
Under Assumption BU, the minimal bid is $\underline{b}=\underline{v}>0$ when at least one bidder attends the auction. Let $G_n (\cdot|d)$ be the bid distribution given $N=n>0$ with support $[\underline{b},\overline{b}_n]$. 

As a convention, the winning bid is set to $0$ when there is no bid. It follows that the winning bid cdf, given that no buyers attend the auction and $N=n$, can be written as $G_n^0 (\cdot|d)$ over $[0,\infty)$. Since the winning bid distribution given $N=n$ and $0<m\leq n$ buyers attend the auction is $G_n^m (\cdot|d)$, the unconditional winning bid distribution is
\begin{align}
G(b|d) & =  \sum_{n=\underline{n}}^{\overline{n}} p_n
\sum_{m=0}^n 
\left(
\begin{array}{c}
m\\
n
\end{array}
\right)
d^{m} (1-d)^{m-n}
G_{n}^{m}\left( b|d\right)
\nonumber \\
& 
=
\sum_{n=\underline{n}}^{\overline{n}} p_n
\left(1-d+d \cdot G_n (b|d) \right)^n.
\label{Mixture_BC}
\end{align}
A key difference with the baseline model is that 
\[
G (\underline{b}|d)
=
\sum_{n=\underline{n}}^{+\overline{n}} p_n
\left(1-d \right)^n
\] 
is now the probability that no buyers attend the auction, which is positive. The lowest number of bidders $\underline{n}$ cannot be identified from the lower tail of $G(\cdot|d)$ as done in the baseline model, where $G(\cdot|d)$ was vanishing and behaving locally as a power $(b-\underline{b})^{\underline{n}}$ near $\underline{b}$.

Some closed-form expressions can be easily obtained for the equilibrium bid quantile function $B_{n} (\cdot|d) =G_{n}^{-1} (\cdot|d)$. Given $N=n$ and assuming that a buyer with private value $V(\alpha)=F^{-1} (\alpha)$ attends the auction, the expected profit generated by a bid $B_{n} (a|d)$ is
\begin{align*}
&\left(V(\alpha)-B_n (a|d)\right) 
\sum_{m=0}^{n-1}
\left(
\begin{array}{c}
m\\
n-1
\end{array}
\right)
d^{m}(1-d)^{n-1-m} a^m
\\
& 
\quad
=
\left(V(\alpha)-B_n (a|d)\right)
\left(1-d+d \cdot a\right)^{n-1}.
\end{align*}
This expression is obtained noting that,
when $n$ bidders are potentially active and at least one is, the distribution of the remaining number of bids is a binomial with parameter $(n-1,d)$. The first-order condition characterizing the Bayesian Nash Equilibrium is then
\[
\frac{d}{d \alpha}
\left[
\left(1-d+d \cdot \alpha\right)^{n-1}
B_n (\alpha|d)
\right]
=
(n-1) d \left(1-d+d \cdot \alpha\right)^{n-2} V(\alpha),
\]
implying, given $B_n(0|d) = \underline{v}$,
\begin{align}
V(\alpha)
&
=
B_n (\alpha|d)
+
\frac{\left(1-d+d \cdot \alpha\right) B_n^{(1)} (\alpha|d)}{(n-1)\cdot d},
\label{Nash_BU}
\\
B_n (\alpha|d)
&
=
\frac{
	(1-d)^{n-1} \underline{v}
	+
	(n-1)\cdot d \cdot
	\int_{0}^{\alpha}
	\left(1-d+d \cdot t \right)^{n-2} V(t)dt
}{\left(1-d+d \cdot \alpha\right)^{n-1}},
\label{Bqf_BU}
\end{align}
which are the counterparts of (\ref{Nash}) and (\ref{Bqf}). In particular, integrating by parts in (\ref{Bqf_BU}) shows that
\[
B_n (\alpha|d)
=
V(\alpha)
-
\int_{0}^{\alpha}
\left( 
\frac{1-d+d \cdot t}{1-d + d \cdot \alpha}
\right)^{n-1}
V^{(1)} (t) dt,
\]
which implies that $B_n (\alpha|d)$ is strictly increasing with respect to $n$ for all quantile levels $\alpha>0$. Hence the largest bids $\overline{b}_n = B_n (1|d)$ satisfy $\overline{b}_{\underline{n}} <  \overline{b}_{\underline{n}+1} < \cdots < \overline{b}_{\overline{n}}<\overline{v}$ as in the baseline model.

The next lemma describes implications of (\ref{Nash_BU}) and (\ref{Bqf_BU}) for the conditional bid pdf $g_n (b|d) = \frac{d}{db} G_n (b|d)$, which parallels Corollary \ref{Extgn}. Henceforth, we shall  consider the additional parameter
\[
\overline{v}^{(1)} =V^{(1)} (1) = \frac{1}{f(\overline{v})}.
\]

\begin{lemma}
	\label{Extgn_BU}
	Suppose Assumptions BU, N, and IPV hold. Then,
	\begin{align}
	\label{Lowergn_BU}
	g_n (b|d) & =
	\left(
	\frac{2f(\underline{v})(1-d)}{(n-1)\cdot d \cdot (b-\underline{b})}
	\right)^{\frac{1}{2}} (1+o(1))
	\text{ when $b \downarrow \underline{b}$,}
	\\
	g_n (\overline{b}_n|d)
	& 
	=
	\frac{1}{(n-1)\cdot d \cdot (\overline{v}-\overline{b}_n)},
	\label{Uppergn_BU}
	\\
	\label{Upperdergn_BU}
	g_n^{(1)} (\overline{b}_{n}|d)
	& =\frac{
		n \cdot d \cdot 
		\left(\overline{v}-\overline{b}_n\right)
		-
		\overline{v}^{(1)}
	}{\left((n-1)\cdot d \right)^2
		\left( \overline{v}-\overline{b}_n\right)^3}.
	\end{align}
\end{lemma}

As the expression of the winning bid distribution (\ref{Mixture_BC}) gives a pdf of
\[
g(b|d)
=
\sum_{n=\underline{n}}^{\overline{n}} n p_n
\cdot d \cdot 
\left(1-d+d \cdot G_n (b|d) \right)^{n-1} g_n (b|d),
\]
(\ref{Uppergn_BU}) shows that $g(\cdot|d)$ exhibits downward jumps $\Delta_n$ at each $\overline{b}_n$, $n=\underline{n},\ldots,\overline{n}$ with 
\[
\Delta_n =\lim_{t\downarrow 0}\left( g(\overline{b}_n-t|d)-g(\overline{b}
_n+t|d) \right)
=
\frac{np_n}{(n-1)(\overline {v}-\overline{b}_{n})},
\]
an expression identical to the jumps of the baseline model (\ref{Jumps}). As (\ref{Bqf_BU}) ensures that $g(\cdot|d)$ is continuous at other points of $(\underline{b},\overline{b}_{\overline{n}})$, this can be used to identify $\overline{n}-\underline{n}$ and 
$p_{n}=\frac{n-1}{n}\Delta_{n}\left( \overline{v}-\overline{b}_{n}\right)$ up to $n$ and $\overline{v}$.

Equation (\ref{Lowergn_BU}) shows that the conditional bid pdf $g_n (\cdot|d)$ and the winning bid pdf $g(\cdot|d)$ both diverge with a  $-\frac{1}{2}$ power rate at the lowest bid $\underline{b}=\underline{v}$. This is intuitively due to buyer uncertainty, as a bidder can win with a very low bid if no others attend the auction. As a consequence, this can be used to check the presence of bidder uncertainty.\footnote{This can be tested using an asymptotically normal estimation of the lower tail-index exponent of the winning bid distribution, such as those reviewed in Beirlant, Herrmann and Teugels (2016). See also Hill and Shneyerov (2013) who also rely on tail-index estimation to test common against private values in an auction context.} On the other hand, it does not allow for the identification of $\underline{n}$ using the lower tail behavior of $G(\cdot|d)$ as simply as in the baseline model.

\paragraph{Discontinuities and identification strategies.}
A possible way to recover the minimal number $\underline{n}$ of potentially active bidders relies on the derivative pdf discontinuities, as permitted by (\ref{Upperdergn_BU}).  The winning bid derivative pdf is
\begin{align*}
g^{(1)}(b|d)
& 
=
\sum_{n=\underline{n}}^{\overline{n}} n p_n
\cdot d \cdot 
\left(1-d+d \cdot G_n (b|d) \right)^{n-2} 
\\
&
\quad\quad\quad
\times
\left[\left(1-d+d \cdot G_n (b|d) \right)g_n^{(1)} (b|d)+(n-1)\cdot d \cdot g_n^{2}(b|d)\right]
,
\end{align*}
which, by (\ref{Bqf_BU}), is continuous over $(\underline{b},\overline{b}_{\overline{n}})$, with the possible exception of the identified $\overline{b}_n$, $n=\underline{n},\ldots,\overline{n}$, where it may exhibit jumps
\begin{align*}
\Delta_{n}^{(1)}
& = 
\lim_{t\downarrow 0}\left( g^{(1)}(\overline{b}_n-t|d)-g^{(1)}(\overline{b}
_n+t|d) \right)
=
n  d  g_n^{(1)} (\overline{b}_n|d)
+
n(n-1)  d^2 g_n^{2} (\overline{b}_n|d)
\\
& =
p_n
\frac{n}{n-1}
\frac{1}{(\overline{v}-\overline{b}_n)^2}
\left(
2+\frac{1}{n-1}
-
\frac{1}{n-1} \frac{\overline{v}^{(1)}}{d} 
\frac{1}{\overline{v}-\overline{b}_n}	
\right),
\end{align*}
by (\ref{Uppergn_BU}), (\ref{Upperdergn_BU}), and a little algebra. To take advantage of the fact that $n-\underline{n}$  is identified as the rank of $\overline{b}_n$,\footnote{This follows from $\overline{b}_{\underline{n}}<\cdots<\overline{b}_{\overline{n}}$, defining the rank of $\overline{b}_{\underline{n}}$ as $0$, the one of $\overline{b}_{\overline{n}}$ being $\overline{n}-\underline{n}$. }, set
\[
m=n-\underline{n},
\quad
\overline{b} (m)
=
\overline{b}_{\underline{n}+m},
\quad
\varrho (m) = \frac{\Delta^{(1)}_{\underline{n}+m}}{\Delta_{\underline{n}+m}},
\]
which are all identified.
It then follows from the expression of $\Delta^{(1)}_{n} $ and $\Delta_{n}$ given the above that
\begin{equation}
\varrho (m)
(\underline{n}+m-1) (\overline{v}-\overline{b} (m))^2
-
\left(2\underline{n}+2m-1\right)(\overline{v}-\overline{b} (m))
-
\frac{\overline{v}^{(1)}}{d}
=
0.
\label{Identeq_BU}
\end{equation}
As this equation includes the three unknowns $\underline{n}$, $\overline{v}$, and $\overline{v}^{(1)}/d$, three of these equations are, in principle, needed for identification from (\ref{Identeq_BU}). However, implementing this strategy may be hard due to the nonlinear nature of (\ref{Identeq_BU}).

A simple overidentification strategy introduces as additional unknowns some well chosen nonlinear functions, such as $\underline{n}
\left(\overline{v}-\overline{b} (0)\right)^2$, 
$
\left(\overline{v}-\overline{b} (0)\right)^2$, and
$\underline{n}
\left(\overline{v}-\overline{b} (0)\right)$,
to transform (\ref{Identeq_BU}) into a linear equation.  This allows us to obtain a condition ensuring identification of the considered auction specification as stated in the next proposition.

\begin{proposition}
	\label{Ident_BU}
	Suppose Assumptions BU, N, and IPV hold, and that $\overline{n}-\underline{n}\geq 5$. Assume there is a subset $\mathcal{M}$ of $\left\{0,\ldots, \overline{n}-\underline{n} \right\}$ with six elements, and let $I_{\mathcal{M}}$ be the $6 \times 6$ matrix with the following row entries, for $m$ in $\mathcal{M}$,  
	\begin{align*}
	\bigg[
	1,
	\text{ }
	\varrho (m),
	\text{ }
	m \varrho (m),
	\text{ }
	\overline{b} (m)
	-
	\overline{b} (0),
	\text{ }
	2m+
	(m-1)
	\varrho (m)
	\left(\overline{b} (m)
	-
	\overline{b} (0)\right)
	,
	&
	\\
	\left(\overline{b} (m)
	-
	\overline{b} (0)\right)
	\left(
	\varrho (m)
	\left(\overline{b} (m)
	-
	\overline{b} (0)\right)
	+
	2
	\right)
	\bigg]
	.&
	\end{align*}
	
	Then, if $\det(I_{\mathcal{M}}) \neq 0$, the uncertainty probability $d$, $\underline{n}$, $\overline{n}$, and $p_n$ for $n=\underline{n},\ldots, \overline{n}$ together with the private value distribution $F(\cdot)$ are identified.
\end{proposition}

As $\varrho (m)$ and $\overline{b}_n$ can be estimated from the data, the overidentification condition $\det(I_{\mathcal{M}}) \neq 0$ is testable. Proposition \ref{Ident_BU} holds under the condition $\overline{n}-\underline{n}\geq 5$, a condition that can be weakened by introducing fewer additional unknowns in (\ref{Identeq_BU}) and by taking into account that $\underline{n}$ is an integer number.\footnote{Note that Proposition \ref{Ident_BU} also applies when buyers are certain about participation (i.e. $d=1$), in which case $\underline{n}$ is identified without relying on the tail argument used in Lemma \ref{Idprelims}-(iii).  \label{Notail}}

\paragraph{Buyer uncertainty and number of buyers.} To some extent, the entry probability $d$ in \eqref{Identeq_BU} can depend upon the number of potential bidders $n$. For instance, the parameters $\beta_0$ and $\beta_1$ in $d_n =1/(\beta_0 + \beta_1 \cdot n)$ can be identified using a straightforward modification of Proposition \ref{Ident_BU}.

Entry cost models as considered in Li and Zheng (2009) also generates buyer uncertainty with a parameter $d$ depending upon $n$ and an entry cost $c$. If in addition the private value distribution depends upon a finite-dimensional $\theta$, then the buyer uncertainty parameter is determined by an indifference condition which ensures $d=d(n,c,\theta)$, assuming $c$ is constant across auctions. Equation \eqref{Identeq_BU} can then be used to identified $c$ and $\theta$, using only that the largest bid $\overline{b}_n$ increases with $n$. Indeed, a difficulty there is that, as noted by Li and Zheng (2009), bidding strategies may not increase with the number of potential bidders. If $c$ varies across auction, nonparametric identification can be restored if the entry cost can be small enough to allow entry of all bidders, suggesting that cost variation can be useful. Indeed, if $d = 1$ for small enough entry costs, identification follows from the baseline case.

\subsection{Unobserved auction heterogeneity \label{UH}}

\paragraph{Setup and assumptions.}
Consider now a setup with auction heterogeneity, where for each buyer $i$,
\begin{equation}
\widetilde{V}_i = \chi + V_i,
\label{UAHeq}
\end{equation}
$\chi$ being an auction-specific variable which is not observed by the econometrician but common knowledge to buyers, and $V_i$ is an i.i.d. private value component drawn from $F(\cdot)$ satisfying Assumption IPV.

\bigskip

\textbf{Assumption UAH}.\textit{\ The unobserved auction heterogeneity component $\chi$ is independent of $N$ and all the private values $V_i$. The pdf $\varphi (\cdot)$ of $\chi$ has a compact support $[0,\overline{\chi}] \subset [0,\infty)$, over which it is strictly positive and twice continuously differentiable. $\overline{\chi} \neq \overline{b}_n - \overline{b}_m$ for all $\underline{n} \leq n,m \leq \overline{n}$.}

\bigskip

The restriction on $\overline{\chi}$ shortens some proofs but can be easily removed.

\paragraph{Model primitives.}
Under (\ref{UAHeq}), the bids $\widetilde{B}_i$ are equal to $\chi + B_i$, where the conditional quantile function $B_n(\cdot)$ of the i.i.d. $B_i$ given $N=n$ is given by (\ref{Bqf}) and satisfies (\ref{Nash}). It follows that the winning bid $\widetilde{W}$ is now
\[
\widetilde{W} = \chi + W, \text{ where } W = \max_{i \in \mathcal{N}} B_i.
\]
The conditional cdf of $W$ is the one of the baseline model, so that, $\Phi (\cdot)$ being the cdf of $\chi$,
\begin{align*}
\widetilde{G}_n^n (b)
& 
=
\mathbb{P}
\left( \left. \widetilde{W} \leq b \right| N=n \right)
=
\mathbb{P}
\left( \left. \chi \leq b - W \right| N=n \right) 
=
\int_{\underline{b}}^{\overline{b}} \Phi (b-t) n G_n^{n-1} (t) g_n (t) dt, 
\end{align*}
as the pdf of $W$ is $n G_n^{n-1} (b) g_n (b)$. It follows that the pdf of $\widetilde{W}$ is, by Assumption N and recalling that $\chi$ belongs to $[0,\overline{\chi}]$, 
\begin{align}
\widetilde{g} (b)
& 
=
\sum_{n=\underline{n}}^{\overline{n}}
p_n
\int_{b-\overline{\chi}}^{b} \varphi (b-t) n G_n^{n-1} (t) g_n (t) dt
\nonumber
\\
&
=
\int_{b-\overline{\chi}}^{b} \varphi (b-t)
g(t)
dt
\text{ where }
g(t)
=
\sum_{n=\underline{n}}^{\overline{n}}
p_nn G_n^{n-1} (t) g_n (t)
,
\label{Tildeg}
\end{align}
noting that $g(\cdot)$ is the winning bid pdf of the baseline model.

\paragraph{Winning bid pdf derivatives discontinuities.}
Integrating out $g(\cdot)$ in (\ref{Tildeg}) gives a smooth pdf $\widetilde{g} (\cdot)$. However discontinuities arise when differentiating $\widetilde{g} (\cdot)$. It indeed holds that applying the Leibniz rule for integral differentiation to (\ref{Tildeg}) yields
\begin{align}
\widetilde{g}^{(1)} (b)
& = 
\sum_{n=\underline{n}}^{\overline{n}}
p_n
n
\left[ \varphi(0) G_n^{n-1} (b) g_n (b)
-
\varphi(\overline{\chi}) G_n^{n-1} (b-\overline{\chi}) g_n (b-\overline{\chi})
\right]
\label{Tildeg1}
\\
& \quad
+
\sum_{n=\underline{n}}^{\overline{n}}
p_n
\int_{b-\overline{\chi}}^{b} \varphi^{(1)} (b-t) n G_n^{n-1} (t) g_n (t) dt.
\nonumber
\end{align}
As the integral expression above remains continuous, (\ref{Tildeg1}) implies that the discontinuities of $\widetilde{g}^{(1)} (\cdot)$  arise at each
$\overline{b}_n$ and $\overline{b}_n+\overline{\chi}$, with jumps that are opposite in sign but of proportional magnitude.\footnote{Hence, if $\overline{b}_n=\overline{b}_m+\overline{\chi}$ for some $n$ and $m$, some upward or downward jumps can be set to $0$ and then washed out. Assumption UAH simplifies identification here by ruling out this possibility.} The next lemma summarizes some properties of $\widetilde{g} (\cdot)$ and its first and second derivatives. Recall that $\overline{v}^{(1)}=V^{(1)}(0)=1/f(\underline{v})$.
\begin{lemma}
	\label{Extgn_UAH}
	Suppose Assumptions BU, N, and IPV hold. Then:
	\begin{enumerate}
		\item $\widetilde{g}(\cdot)$ is continuous over $[0,\infty)$ with $\underline{n}=\lim_{t\downarrow 0}\frac{\log \widetilde{g}(\underline{b}+t)}{\log t}$;
		\item
		$\widetilde{g}^{(1)} (\cdot)$ is continuous over $[0,\infty)$, except at $\overline{b}_n$ and $\overline{b}_n + \overline{\chi}$, $n=\underline{n},\ldots,\overline{n}$. It has a downward jump at $\overline{b}_n$ of size
		\[
		\widetilde{\Delta}_n = \varphi (0) p_n \frac{n}{n-1} \frac{1}{\overline{v}-\overline{b}_n}
		\]
		and an upward jump at $\overline{b}_n+\overline{\chi}$ of size
		$\frac{\varphi (\overline{\chi})}{\varphi (0)}\widetilde{\Delta}_n$;
		\item
		$\widetilde{g}^{(2)} (\cdot)$ is continuous over $(0,\infty)$, except at $\overline{b}_n$ and $\overline{b}_n + \overline{\chi}$, $n=\underline{n},\ldots,\overline{n}$. It has a  jump at each $\overline{b}_n$, of size
		\[
		\widetilde{\Delta}_n^{(1)} = p_n
		\left[
		\varphi(0)
		n
		\frac{(2n-1)\left(\overline{v}-\overline{b}_n\right)- \overline{v}^{(1)}
		}{
			(n-1)^2\left(\overline{v}-\overline{b}_n\right)^3
		}
		+
		\varphi^{(1)} (0)
		\frac{n}{n-1}\frac{1}{\overline{v}-\overline{b}_n}
		\right]
		\]
		and a jump at each $\overline{b}_n+\overline{\chi}$.
	\end{enumerate}
\end{lemma}

\paragraph{Identification of the participation distribution $p_n$.}
Lemma \ref{Extgn_UAH}-(i) ensures that $\underline{n}$ is identified while (ii) implies that the number of jumps of $\widetilde{g}^{(1)} (\cdot)$ above $\underline{b}$ is $2(\overline{n}-\underline{n})$ under the restriction on $\overline{\chi}$ of Assumption AUH, so that $\overline{n}$ can also be recovered. It also shows that $\overline{b}_{\underline{n}}<\cdots<\overline{b}_{\overline{n}}$ are identified as locations of  downward jumps. As upward jumps are located at $\overline{b}_{\underline{n}}+\overline{\chi}<\cdots<\overline{b}_{\overline{n}}+\overline{\chi}$, the unobserved heterogeneity upper bound support  $\overline{\chi}$ is also identified. This may help identify parametric heterogeneity distributions that depend upon a unique parameter.

Identifying the participation distribution is more difficult than in the baseline model because the downward jumps $\widetilde{\Delta}_{n}$ in Lemma \ref{Extgn_UAH}-(ii) now depend upon two unknown parameters, $\overline{v}$ and $\varphi (0)$, unless the latter is identified. Introducing the upward jumps do not help, as they depend on the unknown $\overline{v}$ and $\varphi (\overline{\chi})$. To address this issue, one can try to identify  $\overline{v}$ using the discontinuity ratio $\widetilde{\varrho}_n = \frac{\widetilde{\Delta}^{(1)}_n}{\widetilde{\Delta}_n}$, which, by Lemma \ref{Extgn_UAH}-(ii,iii), satisfies
\begin{equation}
\label{Identeq_AUH}
(n-1)
\left( 
\widetilde{\varrho}_n - \frac{\varphi^{(1)}(0)}{\varphi(0)} 
\right)
\left(\overline{v}-\overline{b}_n\right)^2
-
(2n-1)\left(\overline{v}-\overline{b}_n\right)
-
\overline{v}^{(1)}=0.
\end{equation}

This equation involves three unknowns, $\overline{v}$, $\overline{v}^{(1)}$, and the  ratio $\varphi^{(1)}(0)/\varphi(0)$. Three of these equations are, in principle, needed for identification. Similar to equation (\ref{Identeq_BU}), it can be used as  for (over)identification purposes, considering several values of $n$ and introducing extra variables that are nonlinear functions of the initial unknowns to back out  $\overline{v}$, $\overline{v}^{(1)}$, and  $\varphi^{(1)}(0)/\varphi(0)$ as the unique solution of an extended linear system.

\begin{proposition}
	\label{Ident_AUH}
	Suppose Assumptions BU, N, and IPV hold, and that $\overline{n}-\underline{n} \geq 5 $. Assume there is a subset $\widetilde{\mathcal{N}}$ of $\left\{\underline{n},\ldots, \overline{n} \right\}$ 
	with six elements, and let $I_{\widetilde{\mathcal{N}}}$ be the $6 \times 6$ matrix with the following row entries, for $n$ in $\widetilde{\mathcal{N}}$, 
	\begin{align*}
	\bigg[
	1,
	\text{ }
	n,
	\text{ }
	(n-1)
	\widetilde{\varrho}_n
	\text{ }
	,
	(n-1)
	\left(
	\overline{b}_{n}
	-
	\overline{b}_{\underline{n}}\right),
	\text{ }
	(n-1)
	\left(\overline{b}_{n}
	-
	\overline{b}_{\underline{n}}\right)^2,
	\text{ }
	(n-1)
	\left(
	\overline{b}_{n}
	-
	\overline{b}_{\underline{n}}\right)
	\widetilde{\varrho}_n
	\bigg]
	.&
	\end{align*}
	
	Then, if $\det(I_{\widetilde{\mathcal{N}}}) \neq 0$, the participation distribution $\left\{p_n,n=\underline{n},\ldots,\overline{n}\right\}$ is identified, as $\underline{v}$, $\overline{v}$, and $\overline{v}^{(1)}$.
	
	In addition, $\overline{\chi}$, $\varphi (0)$, $\varphi^{(1)} (0)$, $\varphi (\overline{\chi})$, and  $\varphi^{(1)} (\overline{\chi})$ are also identified.
\end{proposition}

Proposition \ref{Ident_AUH} gives a testable condition, ensuring that the participation distribution can be identified. It also leaves the door open for identification of parametric private value distributions with a parameter in a one-to-one correspondence with $\left(\underline{v}, \overline{v},\overline{v}^{(1)}\right)$, as the latter can be identified. As explained before the Proposition, the participation distribution is identified using only downward jumps at each $\overline{b}_n$. Hence using also upward jumps at $\overline{b}_n+\overline{\chi}$ as well as differences between upward and downward jump locations yield overidentification of the participation distribution and $\overline{\chi}$, which can be used in principle for specification testing.

Similarly, the parametric unobserved heterogeneity distribution that can be uniquely recovered from 
$\left(\overline{\chi}, \varphi (0), \varphi^{(1)} (0), \varphi (\overline{\chi}), \varphi^{(1)} (\overline{\chi}),\right)$ can also be identified. As the expression of the winning bid pdf $\widetilde{g} (\cdot)$ in (\ref{Tildeg}) shows that it is the convolution of the baseline winning bid pdf $g(\cdot)$ by $\varphi (\cdot)$, 
 the deconvolution technique of Krasnokutskaya (2011)  using the identified $\varphi(\cdot)$ allows for the recovery of $g(\cdot)$. If so, the identification procedure developed for the baseline model can be applied to nonparametrically recover the private value distribution.

\section{Final remarks \label{FR}}

This paper shows that, under the independent symmetric private value paradigm,  the first-price winning bid is sufficient to identify model primitives when competition is observed by the buyers but not the econometrician. This is suitable in the presence of phantom bids or when the number of observed bids does not reflect participation. To some extent, buyers can be uncertain about their competitors and auction-specific unobserved heterogeneity can be present. 

Our theoretical results shed light on new identification arguments for discrete mixture models, which are widely used in economic applications, in particular when unobserved heterogeneity is plausible. In our model, the mixture components are generated by the same function. The components are ordered according first-order stochastic dominance and their supports are nested. These two features may appear in other relevant economic mixtures. Investigating how essential these features are for identification can also be of interest for future extensions.

Building on our constructive identification strategy, a natural next step is to develop an estimation procedure. Although the identification approach is constructive, its direct implementation in estimation poses challenges due to the iterative nature of the procedure, which can amplify errors. As discussed in Section 2.6, there is apparently no available methodology that could tackle the estimation problem at hand, even in a parametric framework. In ongoing work, we are developing an estimation framework to address these issues. In particular, we explore the use of Bayesian methods, which offer a coherent way to incorporate prior information, impose structural restrictions, and enhance computational tractability.

 \renewcommand{\thesubsection}{A.\arabic{subsection}}
 \renewcommand{\theproposition}{A.\arabic{proposition}}
 \renewcommand{\thetheorem}{A.\arabic{theorem}}
 \renewcommand{\theclaim}{A.\arabic{claim}}

 \pagebreak
 
 \setcounter{page}{1} 
 
 \renewcommand{\thesubsection}{A.\arabic{subsection}}
 \renewcommand{\theproposition}{A.\arabic{proposition}}
 \renewcommand{\thetheorem}{A.\arabic{theorem}}
 \renewcommand{\theclaim}{A.\arabic{claim}}

\section*{Appendix A: proofs}

\subsection{Proof of Proposition \ref{GIPVNknown}}

It remains to be shown
that (i) and (ii) are sufficient. The function $V(\cdot)$ in (ii) is a
quantile function associated with a cdf $F(\cdot)$ satisfying the
requirements of Assumption IPV, while the mixture weights $p_n$ define a
distribution for $N$, as in Assumption N. These $\{p_n, \underline{n} \leq n
\leq \overline{n} \}$ and private value quantile function $V(\cdot)$
generate a distribution for $N$ and best response bidding strategy functions 
$B_n (\cdot)$ by (\ref{Bqf}), with $G(\cdot)$ as a winning bid cdf.$
\hfill\square$

\subsection{Proof of Corollary \ref{Extgn}}

The compatibility conditions imply that (\ref{Bqf}) holds, and integrating by parts gives
\[
B_{n}\left( \alpha\right) 
=
\frac{1}{\alpha^{n-1}}\int_{0}^{\alpha}
V
\left( t\right) d
\left[ t^{n-1} \right]
=
V\left( \alpha\right) -\int_{0}^{\alpha}\left( \frac{t}{%
	\alpha}\right) ^{n-1}V^{\left( 1\right) }\left( t\right) dt
.  \]
Hence,
$
\overline{b}_n 
=
\overline{v}
-
\int_{0}^{1}t^{n-1}V^{\left( 1\right) }\left( t\right) dt
<\overline{v} $
as $V^{\left( 1\right) }\left( \cdot \right)>0$.
Note that this also gives $B_n (\alpha) < V(\alpha)$ for all $\alpha>0$, and then $B^{(1)} (\alpha)>0$ by (\ref{Nash}). When $\alpha$ goes to $0$, the following holds
\begin{eqnarray*}
	B_n (\alpha)
	& =&
	V(0) + V^{(1)} (0) \alpha + o(\alpha)
	-
	\int_{0}^{\alpha}\left( \frac{t}{%
		\alpha}\right) ^{n-1}
	\left(
	V^{\left( 1\right) }\left( 0\right) 
	+o(1)
	\right)
	dt
	\\
	& = &
	V(0) + \frac{n-1}{n}V^{(1)} (0) \alpha + o(\alpha) , 
\end{eqnarray*}
which shows that $B_n^{(1)} (0) = \frac{n-1}{n}V^{(1)} (0)$.
As $B_n^{(1)} (\cdot)>0$, 
the conditional bid pdf $g_n (\cdot)$ satisfies
\begin{equation}
g_{n}\left( b\right) =\frac{1}{B_{n}^{\left( 1\right) }\left( G_{n}\left(
	b\right) \right) }\text{ for all }b \in \left[ \underline{v},\overline{%
	b}_{n},\right] .
\label{Bn2gn}
\end{equation}
Hence, $g_n (\underline{v}) = 1/B_n^{(1)} (0) = \frac{n}{n-1} 1/V^{(1)} (0) = \frac{n}{n-1} f(\underline{v})$, which is (\ref{Lowergn}). For (\ref{Uppergn}), (\ref{Nash}) and (\ref{Bn2gn}) give
\begin{equation*}
g_{n}\left( \overline{b}_n \right) = 
\frac{G_{n}\left( \overline{b}_n \right) }{\left( n-1\right) \left(
	V\left( G_{n}\left( \overline{b}_n\right) \right) -\overline{b}_n \right) }
=
\frac{1 }{\left( n-1\right) \left(
	\overline{v} -\overline{b}_n \right) }
\end{equation*}
as
$
G_{n}\left( \overline{b}_n \right) = 1
$, so that (\ref{Uppergn}) holds.
$
\hfill\square$

\subsection{Proof of Lemma \ref{Idprelims}} 
As
\[
B_{n}\left( \alpha\right) 
=
\frac{1}{\alpha^{n-1}}
\int_{0}^{\alpha}
V
\left( t\right) d
\left[ t^{n-1} \right]
=
V\left( \alpha\right) -\int_{0}^{\alpha}\left( \frac{t}{%
	\alpha}\right) ^{n-1}V^{\left( 1\right) }\left( t\right) dt , 
\]
differentiating with respect to $n$ gives 
\begin{eqnarray*}
	\frac{\partial B_{n}\left( \alpha\right) }{\partial n}
	&=&
	-\int_{0}^{\alpha}\left( \frac{t}{\alpha}\right) ^{n-1}\log\left( \frac{t}{%
		\alpha}\right) V^{\left( 1\right) }\left( t\right) dt\geq 0 .
\end{eqnarray*}
The inequality is strict except when $\alpha=0$, in which case $%
B_{n}\left( 0\right) =\underline{v} $ for all $n$. It follows that the bid
cdf given that $N=n$, $G_n (\cdot)$, has a support $[\underline{v},%
\overline{b}_n]$, with an upper bound $\overline{b}_n=B_n (1)$, which is
strictly increasing with respect to $n$ and strictly smaller than $\overline{	v}=\lim_{n \uparrow \infty} \overline{b}_n$. Hence, this proves Part (i). For part (ii), the expression for jumps (\ref{Jumps}) follows from (\ref{GNknown}), which shows that the winning bid pdf is
\begin{equation}
g\left( b\right) =\sum_{k=\underline{n}}^{\overline{n}} p_{k} k
G_{k}^{k-1}\left( b\right) g_{k}\left( b\right), \label{Winpdf}
\end{equation}
with $g_k (b)=0$ for $b>\overline{b}_n$ when $k \leq n$ by Lemma \ref{Idprelims}-(i). This gives 
\begin{eqnarray*}
	& &g (\overline{b}_n-t)-g (\overline{b}_n+t)
	\\
	& &\quad = 
	\sum_{k=n}^{\overline{n}} p_{k} k
	G_{k}^{k-1}\left( \overline{b}_n-t\right) g_{k}\left( \overline{b}_n-t\right)
	-
	\sum_{k=n+1}^{\overline{n}} p_{k} k
	G_{k}^{k-1}\left( \overline{b}_n+t\right) g_{k}\left( \overline{b}_n+t\right)
	\\
	& &\quad 
	\rightarrow
	n p_n g_n \left( \overline{b}_n \right)
	=
	\Delta_n , 
\end{eqnarray*}
when $t$ goes to $0$.
The equality (\ref{Uppergn}) for $g_n \left( \overline{b}_n \right)$ then gives (\ref{Jumps}). For part (iii), continuity of $B_n^{(1)}(\cdot)$, which is bounded away from $0$ and infinity, and (\ref{Bn2gn}) shows that $g_n (\cdot)$ is continuous with $g_n(\underline{v})>0$ by (\ref{Lowergn}). When $t$ goes to $0$, this gives 
\begin{align*}
G \left(\underline{v} + t \right)
& = 
\sum_{n=\underline{n}}^{\overline{n}}
p_n
\left(
\int_{\underline{v}}^{\underline{v}+t}
g_n (u) du
\right)^n
=
\sum_{n=\underline{n}}^{\overline{n}}
p_n
g_n^n (\underline{v}) t^n
\left(1+o(1)\right)
\\
& = 
p_{\underline{n}} g_{\underline{n}}^{\underline{n}} (\underline{v}) t^{\underline{n}}
\left(1+o(1)\right) , 
\end{align*} 
as $p_{\underline{n}} g_{\underline{n}}^{\underline{n}} (\underline{v})>0$, which implies $\underline{n}=\lim_{t\downarrow0}\frac{\log G\left( 
	\underline{v}+t\right) }{\log t}$.
$\hfill\square$

\subsection{Proof of Theorem \ref{FP}}

We now obtain identification of the value quantile function by iteratively exploiting the two equilibrium mappings in (\ref{Nash}) and (\ref{Nash2}). We proceed in three steps: 

\subparagraph{Step 1.}

Note that the winning bid distribution satisfies 
\begin{equation*}
G\left( b\right) =1-p_{\overline{n}}+p_{\overline{n}}G_{\overline{n}}^{%
	\overline{n}}\left( b\right) \text{ for all }b\text{ in }\left[ \overline{b}%
_{\overline{n}-1},\overline{b}_{\overline{n}}\right]
\end{equation*}
so that $G_{\overline{n}}\left( \cdot \right)$ is identified over $\left[ 
\overline{b}_{\overline{n}-1},\overline{b}_{\overline{n}}\right]$ as follows: 
\begin{equation*}
G_{\overline{n}}\left( b\right) =\left( \frac{G\left( b\right) -\left( 1-p_{%
		\overline{n}}\right) }{p_{\overline{n}}}\right) ^{1/\overline{n}}\text{ for }%
b\text{ in }\left[ \overline{b}_{\overline{n}-1},\overline{b}_{\overline{n}}%
\right] . 
\end{equation*}
Set 
\begin{equation*}
\alpha_{1}=G_{\overline{n}}\left( \overline{b}_{\overline{n}-1}\right) .
\end{equation*}
It follows that $B_{\overline{n}}\left( \cdot\right)$ is identified on $%
[\alpha_1,1]$, i.e., 
\begin{equation*}
B_{\overline{n}}\left( \alpha\right) =W\left[ \left( 1-p_{\overline{n}%
}\right) +p_{\overline{n}}\alpha^{\overline{n}}\right] ,
\end{equation*}
where $W\left( \cdot\right) =G^{-1}\left( \cdot\right) $ is the winning bid
quantile function. 

Using the mapping from the bid quantile function to the value quantile function (\ref{Nash}) shows that the private value quantile function
satisfies, for all $\alpha \in \left[ \alpha_{1},1\right] $,%
\begin{align*}
V\left( \alpha\right) & = B_{\overline{n}}\left( \alpha\right) + \frac {1}{%
	\overline{n}-1}\alpha B_{\overline{n}}^{(1)}\left( \alpha\right) \\
& = W \left[ \left( 1-p_{\overline{n}}\right) +p_{\overline{n}}\alpha^{%
	\overline{n}} \right] + \frac{\overline{n} p_{\overline{n}} }{\overline{n}-1}
\alpha^{\overline{n}} W^{\left( 1\right) }\left[ \left( 1-p_{\overline{n}%
}\right) +p_{\overline{n}}\alpha^{\overline{n}}\right] , 
\end{align*}
and $V(\cdot)$ is identified over $[\alpha_1,1]$.

Using the mapping from the value quantile function to the bid quantile function (\ref{Nash2}) shows that the bid quantile functions $B_{n}\left( \cdot\right) $, $n=\underline{n},\ldots,\overline{n}-1$ are also identified over $[\alpha_1,1]$.  Hence, $%
\left\lbrace B_{n}(\alpha), \alpha \in [\alpha_1,1] \right\rbrace $ and $%
\left\lbrace G_{n} (b), b \in [B_{n}(\alpha_{1}),\overline{b}%
_{n}]\right\rbrace $ are identified,  for all $n=\underline{n},\ldots,\overline{n}$.

\subparagraph{Step 2.}

We now expand the interval over which $G_{\overline{n}}(\cdot)$ is identified
using an iterative argument. Define%
\begin{equation*}
\beta_{1}=B_{\overline{n}-1}\left( \alpha_{1}\right) , 
\end{equation*}
which is identified from the last step. Note that $\beta_{1}<\overline {b}_{%
	\overline{n}-1}$ whenever $\alpha_{1}>0$ because, by Lemma \ref{Idprelims}%
-(i), 
\begin{equation*}
\beta_{1}=B_{\overline{n}-1}\left[ G_{\overline{n}}\left( \overline {b}_{%
	\overline{n}-1}\right) \right] <B_{\overline{n}}\left[ G_{\overline {n}%
}\left( \overline{b}_{\overline{n}-1}\right) \right] =\overline {b}_{%
	\overline{n}-1}.
\end{equation*}

The definition of $G(\cdot)$ implies that 
\begin{equation}
G_{\overline{n}}\left( b\right) =\left( \frac{G\left(
	b\right) -\sum_{n=\underline{n}}^{\overline{n}-1}p_{n}G_{n}^n\left( b\right)}{p_{\overline{n}}} \right)^{1/\overline{n}} ,  \label{Gtopid}
\end{equation}
where $G\left( \cdot\right) $ and $p_{n}$ are identified, and $G_{n}(\cdot) $
are identified on  $[B_{n}(\alpha_{1}),\overline{b}_{\overline{n}}]$ for all $n=%
\underline{n},\ldots,\overline{n}-1$. Since $B_{\underline{n}}(\alpha
_{1})<\ldots<B_{\overline{n}-1}(\alpha_{1})=\beta_{1}$, $\lbrack\beta_{1},\overline{b}_{\overline{n}}] \subseteq [B_{n}(\alpha _{1}),
\overline{b}_{\overline{n}}]$ for all $n$. Therefore, the conditional bid
distribution $G_{\overline{n}}\left( b\right) $ is identified on $[\beta
_{1},\overline{b}_{\overline{n}}]$.

\subparagraph{Step 3.}

We now identify $V(\cdot)$ on a growing interval $[\alpha_{k},1]$ using an
induction argument and the identified $V(\cdot)$ on $\left[ \alpha _{1},1%
\right] $. For an integer $k\geq2$, define 
\begin{equation*}
\alpha_{k}=G_{\overline{n}}\left( \beta_{k-1}\right) =G_{\overline{n}}\left[
B_{\overline{n}-1}\left( \alpha_{k-1}\right) \right] ,\quad \beta_{k}=B_{%
	\overline{n}-1}\left( \alpha_{k}\right) .
\end{equation*}
Identification of $V(\cdot)$ on the growing interval $[\alpha_{k},1]$ is
established in Lemma \ref{Induction} below.

\begin{lemma}
	\label{Induction} Suppose Assumptions N and IPV hold. Then,
	
	\begin{enumerate}
		\item the sequences $\left\{ \alpha_{k},k\geq1 \right\} $ and $\left\{
		\beta_{k},k\geq1\right\} $ are decreasing sequences with 
		\begin{equation*}
		\lim_{k\rightarrow\infty}\alpha_{k}=0.
		\end{equation*}
		
		\item $\left\{ \alpha_{k},k\geq1 \right\} $ is identified. For
		any integer number $k\geq2$, $\left\{ V\left( \alpha\right) ,\alpha \in\left[
		\alpha_{k},1\right] \right\} $ is identified if $\left\{ V\left(
		\alpha\right) ,\alpha\in\left[ \alpha_{k-1},1\right] \right\} $ is
		identified.
	\end{enumerate}
\end{lemma}

The proof of Lemma \ref{Induction} is given at the end of this section.
Let us now return to the identification of $V(\alpha)$ for any arbitrary $\alpha>0$. By Lemma \ref%
{Induction}-(i), there exists $k$ such that $\alpha>\alpha_{k}$ and Lemma %
\ref{Induction}-(ii) yields identification of $V\left( \alpha\right) $. Given that $%
V\left( 0\right) =\underline{v}$ is identified by Lemma \ref{Discontinuity},
the theorem is proven.$\hfill\square$

\paragraph{Proof of Lemma \protect\ref{Induction}.}
Consider (i) first. As $\alpha_{k}=G_{\overline{n}}\left[ B_{\overline {n}%
-1}\left( \alpha_{k-1}\right) \right] $ with $B_{\overline{n}-1}\left(
\alpha\right) \leq B_{\overline{n}}\left( \alpha\right) $, 
\begin{equation*}
\alpha_{k} =G_{\overline{n}}\left[ B_{\overline{n}-1}\left( \alpha
_{k-1}\right) \right] \leq G_{\overline{n}}\left[ B_{\overline{n}}\left(
\alpha_{k-1}\right) \right] =\alpha_{k-1} ,
\end{equation*}
which implies that $\alpha_{k}$ decreases. Moreover, $\beta_{k}=B_{\overline {n}-1}\left( \alpha_{k}\right) $ decreases because $B_{\overline{n}-1}\left( \cdot\right) $ is strictly increasing. Since $\alpha_{k}\geq0$, $\alpha_{k}$ converges to a limit $\alpha$, which satisfies $\alpha =G_{\overline{n}}\left[ B_{\overline{n}-1}\left( \alpha\right) \right] $ under
Assumption IPV. In other words, the limit $\alpha$ satisfies $B_{\overline{n}}\left( \alpha\right) =B_{\overline{n}-1}\left( \alpha\right) $. This gives $\alpha=0$ as $B_{\overline{n}}\left( \alpha\right) >B_{\overline{n}-1}\left(
\alpha\right) $, except for $\alpha=0$.

Now, consider (ii). That $\alpha_{k}$ is identified for all $k$ follows from
an induction argument, observing $\alpha_{1}$ is identified. Suppose then
that $\alpha_{k}$ and $\left\{ V\left( \alpha\right) ,\alpha\in\left[
\alpha_{k},1\right] \right\} $ are identified. Recall
\begin{equation*}
\alpha_{k+1}=G_{\overline{n}}\left( \beta_{k}\right) =G_{\overline{n}}\left[
B_{\overline{n}-1}\left( \alpha_{k}\right) \right] ,\quad \beta_{k+1}=B_{%
\overline{n}-1}\left( \alpha_{k+1}\right) .
\end{equation*}
Then, (\ref{Nash2}) and Lemma \ref{Discontinuity} give that $\left\{ B_{n}\left( \alpha\right)
;\alpha\in\left[ \alpha_{k},1\right] \right\} $, for all $n=\underline{n},\ldots,%
\overline{n}-1$ are identified, as $\beta_k$. Now  (\ref%
{Gtopid}) and Lemma \ref{Discontinuity} show that $G_{\overline{n}}\left( b\right) $ is identified for
all $b\geq\beta_{k}$, and then $\alpha_{k+1}=G_{\overline{n}}\left(
\beta_{k}\right) $ is identified. (\ref{Nash}) then gives
that $\left\{ V\left( \alpha\right) ;\alpha\in\left[ \alpha_{k+1},1\right]
\right\} $ is identified. This ends the proof of the lemma.$\hfill\square$

\subsection{Proof of Lemma \ref{Extgn_BU} \label{ExtgnBU}}

Consider (\ref{Lowergn_BU}) first. Note that (\ref{Bqf_BU}) shows that $B_n(\cdot|d)$ is continuously differentiable over $[0,1]$. Expanding (\ref{Nash_BU}) gives, when $\alpha$ goes to $0$,
\begin{align*}
B_n^{(1)} (\alpha|d)
& 
=
(n-1)\cdot d
\frac{V(\alpha)-B_n(\alpha|d)}{1-d+d \cdot \alpha}
=
(n-1)\cdot d \cdot \alpha
\frac{V^{(1)}(0)-B_n^{(1)}(0|d)}{1-d}
+ 
o (\alpha) ,	
\end{align*}
which implies $B^{(1)} (0|d) = 0$ and then
\begin{align*}
B_n^{(1)} (\alpha|d)
& 
=
\frac{(n-1)dV^{(1)}(0)}{1-d} \alpha + o(\alpha)
=
\frac{(n-1)d}{(1-d) f(\underline{v})} \alpha + o(\alpha),\\
B_n (\alpha)
& =
\underline{b}
+
\frac{(n-1)d}{(1-d) f(\underline{v})} \frac{\alpha^2}{2} + o(\alpha^2) \text{ so that when $b\downarrow \underline{b}$}
\\
G_n (b|d)
& =
\left(
\frac{2f(\underline{v})(1-d)}{(n-1)d}
\left(b -\underline{b}\right)
\right)^{\frac{1}{2}} (1+o(1)).
\end{align*}
This gives (\ref{Lowergn_BU}), noting
\[
g_n(b|d) = \frac{1}{B_n^{(1)} \left(\left.G_n(b|d)\right|d\right)}
=
\left(
\frac{2f(\underline{v})(1-d)}{(n-1)d(b-\underline{b})}
\right)^{\frac{1}{2}} (1+o(1)).
\]

(\ref{Uppergn_BU}) also follows from $g_n(b|d) = 1/B_n^{(1)} \left(\left.G_n(b|d)\right|d\right)$ and (\ref{Nash_BU}), which gives
\begin{align*}
g_n (\overline{b}_n|d)
& 
=
\left.
\frac{1}{(n-1)d}
\frac{1-d+d \cdot G_n(b|d)}{V\left(G_n(b|d)\right)-b}
\right|_{b=\overline{b}_n}
=
\frac{1}{(n-1)d(\overline{v}-\overline{b}_n)}.
\end{align*}

For (\ref{Upperdergn_BU}), first observe that
\begin{align*}
g_n^{(1)} (b|d)
& =
\frac{d}{db}
\left[
\frac{1}{B_n^{(1)}\left(\left. G_n (b|d) \right|d\right)}
\right]
=
-
\frac{
	B_n^{(2)}\left(\left. G_n (b|d) \right| d\right) g_n(b|d)
}{
	\left( B_n^{(1)}\left(\left. G_n (b|d)\right)\right| d\right)^{2}}
\\
& =
-
\frac{
	B_n^{(2)}\left(\left. G_n (b|d) \right| d\right) 
}{
	\left( B_n^{(1)}\left(\left. G_n (b|d)\right)\right| d\right)^{3}},
\end{align*}
where
\begin{align*}
B_n^{(2)} (\alpha|d)
& =
\frac{d}{d \alpha}
\left[
(n-1)\cdot d
\frac{V(\alpha)-B_n(\alpha|d))}{1-d+d \cdot \alpha}
\right]
\\
& =
-(n-1)\cdot d^2
\frac{V(\alpha)-B_n(\alpha|d)}{\left(1-d+d \cdot \alpha\right)^2}
+
(n-1)\cdot d
\frac{V^{(1)}(\alpha)-B_n^{(1)}(\alpha|d)}{1-d+d \cdot \alpha}
\\
& =
-
\frac{n(n-1) \cdot d^2\left(V(\alpha)-B_n(\alpha|d)\right)}{\left(1-d+d \cdot \alpha\right)^2}
+
(n-1)\cdot d
\frac{V^{(1)}(\alpha)}{1-d+d \cdot \alpha}.
\end{align*}
Hence, 
\begin{align*}
g_n^{(1)} (\overline{b}_{n}|d)
& =
-
\frac{B_n^{(2)}(1|d)}{\left(B_n^{(1)}(1|d)\right)^3}
=
\frac{
	(n-1)\cdot d \cdot
	\left[ n \cdot d \cdot \left(\overline{v}-\overline{b}_n\right)
	-\overline{v}^{(1)}
	\right]
}{\left((n-1)\cdot d \cdot (\overline{v}-\overline{b}_n)\right)^3}
\\
& =
\frac{
	n \cdot d \cdot 
	\left(\overline{v}-\overline{b}_n\right)
	-
	\overline{v}^{(1)}
}{\left((n-1)\cdot d \right)^2
	\left( \overline{v}-\overline{b}_n\right)^3}.
\end{align*}
This ends the proof of the lemma.
\hfill $\Box$

\subsection{Proof of Proposition \ref{Ident_BU} \label{IdentBU}}

Set $x_1=\overline{v}-\overline{b} (0)$, $x_2=\underline{n}$, and $x_3=\frac{\overline{v}^{(1)}}{d}$. Additionally, define the extra unknowns
\begin{align*}
& x_4 = \underline{n} \left(\overline{v}-\overline{b} (0)\right)^2,
& & x_5 = \left(\overline{v}-\overline{b} (0)\right)^2,
& & x_6 = \underline{n}\left(\overline{v}-\overline{b} (0)\right),
\end{align*}
and set $y_m =-(m-1)\varrho (m) (\overline{b} (m)-\overline{b} (0))^2 -
(2m-1)
(\overline{b} (m)-\overline{b} (0))$. 

It follows that
\begin{align*}
&
(\underline{n}+m-1) (\overline{v}-\overline{b} (m))^2
=
(\underline{n}+m-1) 
\left(
\overline{v}-\overline{b} (0)
-
(\overline{b} (m)-\overline{b} (0))\right)^2
\\
& 
=
\underline{n} \left(\overline{v}-\overline{b} (0)\right)^2
+
\left(\underline{n} + m-1 \right)
(\overline{b} (m)-\overline{b} (0))^2
\\
&
\quad
+
(m-1)
\left(\overline{v}-\overline{b} (0)\right)^2
-
2(\overline{b} (m)-\overline{b} (0))
\left[
\underline{n}
\left(\overline{v}-\overline{b} (0)\right)
+
(m-1) \left(\overline{v}-\overline{b} (0)\right)
\right]
\\
&
=
-2
(\overline{b} (m)-\overline{b} (0))
(m-1)
\cdot x_1
+
(\overline{b} (m)-\overline{b} (0))^2
\cdot x_2
+
x_4
+
(m-1) \cdot x_5
\\
&
\quad
-2(\overline{b} (m)-\overline{b} (0)) 
\cdot
x_6
+
(m-1)
(\overline{b} (m)-\overline{b} (0))^2 
,
\\
&\left(2\underline{n}+2m-1\right)(\overline{v}-\overline{b} (m))
=
\left(2\underline{n}+2m-1\right)
\left(\overline{v}-\overline{b} (0)
-(\overline{b} (m)-\overline{b} (0))\right)
\\
&
\quad
=
(2m-1) \cdot x_1
-2(\overline{b} (m)-\overline{b} (0)) \cdot x_2
+
2 \cdot x_6
-
(2m-1)
(\overline{b} (m)-\overline{b} (0)).
\end{align*}
Hence, using these new notations shows that (\ref{Identeq_BU}) is equivalent to
\begin{align*}
-
\left(
2m-1 
+ 
(m-1)
\varrho (m)
(\overline{b} (m)-\overline{b} (0)) 
\right)
\cdot
x_1
&
\\
+
(\overline{b} (m)-\overline{b} (0))
\left(
\varrho (m)
(\overline{b} (m)-\overline{b} (0))
+
2
\right)
\cdot x_2
&
\\
-
x_3 
+
\varrho (m) \cdot x_4
+
(m-1) \varrho (m) \cdot x_5
-
2\left(\overline{b} (m)-\overline{b} (0)+1\right)
\cdot
x_6
=
y_m.
&
\end{align*}
Elementary determinant algebra shows that, when $\det(I_{\mathcal{M}}) \neq 0$, the corresponding linear system obtained for $m$ varying across $\mathcal{M}$ uniquely determines $x_1,\ldots,x_6$, and then $\underline{n}$, $\overline{v}$.

Hence $\overline{n}$ is identified using the number of pdf discontinuities. $p_{n}=\frac{n-1}{n}\Delta_{n}\left( \overline{v}-\overline{b}_{n}\right)$ then identifies $p_n$. As $\underline{b}$ and $G(\underline{b}|d)=
\sum_{n=\underline{n}}^{+\overline{n}} p_n
\left(1-d \right)^n$ are identified, $d$ is identified, since $\sum_{n=\underline{n}}^{+\overline{n}} p_n
x^n$ is an identified polynomial function which is strictly increasing in $x$ over $[0,1]$.

Identification of $F(\cdot)$ can then be established as in the baseline model, using  that $B_n (\alpha|d)$ strictly increases with $n$ for $\alpha$ in $(0,1]$ with $B_n (0|d)=\underline{v}$ for all $n$.
\hfill $\Box$

\subsection{Proof of Lemma \ref{Extgn_UAH} \label{ExtgnUAH}}
Continuity of $\widetilde{g}(\cdot)$ follows from (\ref{Tildeg}).
As $g_n(0)>0$ by (\ref{Lowergn}), (\ref{Tildeg}) implies, for $b$ sufficiently close to $\underline{b}$,
\begin{align*}
\widetilde{g} (b)
& 
= 
\int_{\underline{b}}^{b} 
\left(\varphi(0)+o(1)\right)
\left(
\sum_{n=\underline{n}}^{n=\overline{n}}
g_n^{n} (0) (t-\underline{b})^{n-1} (1+o(1))
\right)
dt
\\
&=
\varphi (0) 
g_{\underline{n}}^{\underline{n}} (0)
(b-\underline{b})^{\underline{n}} (1+o(1)),
\end{align*}
which implies $\underline{n}=\lim_{t\downarrow 0}\frac{\log \widetilde{g}(\underline{b}+t)}{\log t}$.

(ii) follows from (\ref{Tildeg1}) and (\ref{Uppergn}), which states that $g_n(\overline{b}_n) = 1/\left((n-1)(\overline{v}-\overline{b}_n)\right)$.
For (iii), differentiating (\ref{Tildeg1}) gives
\begin{align*}
\widetilde{g}^{(2)} (b)
& = 
\sum_{n=\underline{n}}^{\overline{n}}
p_n
n
\left[ \varphi(0) G_n^{n-1} (b) g_n^{(1)} (b)
-
\varphi(\overline{\chi}) G_n^{n-1} (b-\overline{\chi}) g_n^{(1)} (b-\overline{\chi})
\right]
\\
& +
\sum_{n=\underline{n}}^{\overline{n}}
p_n
n(n-1)
\left[ \varphi(0) G_n^{n-2} (b) g_n^{2} (b)
-
\varphi(\overline{\chi}) G_n^{n-2} (b-\overline{\chi}) g_n^{2} (b-\overline{\chi})
\right]
\\
& + 
\sum_{n=\underline{n}}^{\overline{n}}
p_n
n
\left[ \varphi^{(1)}(0) G_n^{n-1} (b) g_n (b)
-
\varphi^{(1)}(\overline{\chi}) G_n^{n-1} (b-\overline{\chi}) g_n (b-\overline{\chi})
\right]
\\
& \quad
+
\sum_{n=\underline{n}}^{\overline{n}}
p_n
\int_{b-\overline{\chi}}^{b} \varphi^{(2)} (b-t) n G_n^{n-1} (t) g_n (t) dt.
\end{align*}
Hence,
\[
\widetilde{\Delta}_n^{(1)}
=
p_n
\left[
n \varphi(0)
\left(
g_n^{(1)} (\overline{b}_n)
+
(n-1)
g_n^{2} (\overline{b}_n)
\right)
+
n
\varphi^{(1)} (0)
g_n (\overline{b}_n)
\right]
.
\]
Let us now compute $g_n^{(1)} (\overline{b}_n)
=
-
B_n^{(2)}(1)/\left( B_n^{(1)} (1) \right)^3$.
(\ref{Nash}) implies
\begin{align*}
B_n^{(2)} (\alpha)
& =
- n(n-1)
\frac{ \left(V(\alpha)-B_n (\alpha)\right)}{\alpha^2}
+
(n-1)
\frac{V^{(1)}(\alpha)}{\alpha}
\end{align*}
so that
\[
B_n^{(2)} (1)=-n(n-1) \left(\overline{v}-\overline{b}_n\right)
+
(n-1) \overline{v}^{(1)}.
\]
Hence, 
\[
g_n^{(1)} (\overline{b}_n)
=
-
\frac{B_n^{(2)}(1)}{\left( B_n^{(1)} (1) \right)^3}
=
\frac{n\left(\overline{v}-\overline{b}_n\right)-\overline{v}^{(1)}
}{
	(n-1)^2\left(\overline{v}-\overline{b}_n\right)^3
}
\]
by computations similar to the ones at the end of Section \ref{ExtgnBU}. This gives
\begin{align*}
\widetilde{\Delta}_n^{(1)}
& = 
\varphi (0)
p_n
\left(
n
\frac{n\left(\overline{v}-\overline{b}_n\right)-\overline{v}^{(1)}
}{
	(n-1)^2\left(\overline{v}-\overline{b}_n\right)^3
}
+
n(n-1)
\left(\frac{1}{(n-1)\left(\overline{v}-\overline{b}_n\right)}\right)^2
\right)
\\
&
\quad
+
\varphi^{(1)} (0)
p_n
\frac{n}{n-1}\frac{1}{\overline{v}-\overline{b}_n}
\\
& =
p_n
\left[
\varphi(0)
\frac{n(2n-1)\left(\overline{v}-\overline{b}_n\right)-n\overline{v}^{(1)}
}{
	(n-1)^2\left(\overline{v}-\overline{b}_n\right)^3
}
+
\varphi^{(1)} (0)
\frac{n}{n-1}\frac{1}{\overline{v}-\overline{b}_n}
\right].
\end{align*}
This ends the proof of the lemma.
\hfill$\Box$

\subsection{Proof of Proposition  \ref{Ident_AUH} \label{IdentAUH}}
Set $x_1=\overline{v}-\overline{b}_{\underline{n}}$, $x_2=\frac{\varphi^{(1)}(0)}{\varphi(0)}$, $x_3 = \overline{v}^{(1)}$  and
\begin{align*}
& &
x_4 = \left(\overline{v}-\overline{b}_{\underline{n}}\right)^2,
& & 
x_5 =  
\frac{\varphi^{(1)}(0)}{\varphi(0)} 
\left(\overline{v}-\overline{b}_{\underline{n}}\right)^2,
& &
x_6 = 
\frac{\varphi^{(1)}(0)}{\varphi(0)} 
\left(\overline{v}-\overline{b}_{\underline{n}}\right)
.
\end{align*}
Set $y_n =-(n-1)
\widetilde{\varrho}_n
(
\overline{b}_{n}
-
\overline{b}_{\underline{n}}
)^2-(2n-1) (\overline{b}_{n}-\overline{b}_{\underline{n}})$. 

As
\begin{align*}
&
(n-1)
\left( 
\widetilde{\varrho}_n - \frac{\varphi^{(1)}(0)}{\varphi(0)} 
\right)
\left(\overline{v}-\overline{b}_n\right)^2
=
(n-1)
\left( 
\widetilde{\varrho}_n - \frac{\varphi^{(1)}(0)}{\varphi(0)} 
\right)
\left(\overline{v}-\overline{b}_{\underline{n}}
-
(
\overline{b}_{n}
-
\overline{b}_{\underline{n}}
)
\right)^2
\\
& \quad
=
(n-1)
\left( 
\widetilde{\varrho}_n - \frac{\varphi^{(1)}(0)}{\varphi(0)} 
\right)
\left(\overline{v}-\overline{b}_{\underline{n}}\right)^2
-
2
(n-1)
(
\overline{b}_{n}
-
\overline{b}_{\underline{n}}
)
\left( 
\widetilde{\varrho}_n - \frac{\varphi^{(1)}(0)}{\varphi(0)} 
\right)
\left(\overline{v}-\overline{b}_{\underline{n}}\right)
\\
&
\quad\quad
-
(n-1)
(
\overline{b}_{n}
-
\overline{b}_{\underline{n}}
)^2
\frac{\varphi^{(1)}(0)}{\varphi(0)}
+
(n-1)
\widetilde{\varrho}_n
(
\overline{b}_{n}
-
\overline{b}_{\underline{n}}
)^2
\\
& \quad
= 
-
2
(n-1)
(
\overline{b}_{n}
-
\overline{b}_{\underline{n}}
)
\widetilde{\varrho}_n
\cdot x_1
-
(n-1)
(
\overline{b}_{n}
-
\overline{b}_{\underline{n}}
)^2
\cdot x_2
+
(n-1) \widetilde{\varrho}_n
\cdot x_4
\\
&
\quad\quad
-
(n-1)
\cdot x_5
+
2(n-1)(
\overline{b}_{n}
-
\overline{b}_{\underline{n}}
)
\cdot x_6
+
(n-1)
\widetilde{\varrho}_n
(
\overline{b}_{n}
-
\overline{b}_{\underline{n}}
)^2,
\\
&-(2n-1) \left(\overline{v}-\overline{b}_{n}\right)
=
-(2n-1) \cdot x_1 + (2n-1) (\overline{b}_{n}-\overline{b}_{\underline{n}}),
\end{align*}
(\ref{Identeq_AUH}) is equivalent to
\begin{align*}
-
\left[
2
(n-1)
(
\overline{b}_{n}
-
\overline{b}_{\underline{n}}
)
\widetilde{\varrho}_n
+
(2n-1)
\right]
\cdot x_1
-
(n-1)
(
\overline{b}_{n}
-
\overline{b}_{\underline{n}}
)^2
\cdot x_2
+
x_3
&
\\
(n-1) \widetilde{\varrho}_n
\cdot x_4
-
(n-1)
\cdot x_5
+
2(n-1)(
\overline{b}_{n}
-
\overline{b}_{\underline{n}}
)
\cdot x_6
=
y_n.
&
\end{align*}
Stacking these equations for $n$ in $\widetilde{M}$ gives a linear system with a unique solution when $\det(I_{\widetilde{\mathcal{M}}}) \neq 0$. Hence, the initial parameters are $\left(\overline{v},\overline{v}^{(1)},\frac{\varphi^{(1)}(0)}{\varphi(0)}\right)$. 
The identity
$\widetilde{\Delta}_n = \varphi (0) p_n \frac{n}{n-1} \frac{1}{\overline{v}-\overline{b}_n}$ in Lemma \ref{Extgn_UAH} and $\sum_{n=\underline{n}}^{\overline{n}} p_n =1$ allow for the recovery of $\varphi (0)$ and $p_n$, $n = \underline{n},\ldots,\overline{n}$. $\overline{\chi}$ has already been identified, and $\varphi (\overline{\chi})$ can be recovered from the upward jump size $\widetilde{\Delta}_n = \varphi (\overline{\chi}) p_n \frac{n}{n-1} \frac{1}{\overline{v}-\overline{b}_n}$. As $\varphi^{(1)} (\overline{\chi})/\varphi (\overline{\chi})$ satisfies an equation similar to (\ref{Identeq_AUH}), $\varphi^{(1)} (\overline{\chi})$ is also identified.
\hfill $\Box$


\end{document}